\documentclass[lettersize,journal]{IEEEtran}
\usepackage{amsmath,amsfonts}
\usepackage{algorithmic}
\usepackage{algorithm}
\usepackage{array}
\usepackage{textcomp}
\usepackage{stfloats}
\usepackage{url}
\usepackage{verbatim}
\usepackage{graphicx}
\usepackage{cite}
\usepackage{xcolor}
\usepackage{multirow}
\usepackage{booktabs}
\usepackage{lipsum}
\usepackage{comment}
\usepackage{ifthen}
\usepackage[hidelinks]{hyperref}

\usepackage{LMSnodeshapes,LMScolor}
\usepackage{subcaption,caption}
\usepackage{pgfplots}
\usepackage[shortcuts,acronym]{glossaries}

\usepackage{tikz}
\usetikzlibrary{calc}
\usetikzlibrary{positioning}
\usetikzlibrary{backgrounds}
\usetikzlibrary{shapes}
\usetikzlibrary{arrows.meta}
\tikzset{>=stealth} 

\newboolean{ieeeSubmission}
\setboolean{ieeeSubmission}{false}

\begin{document}

\newacronym{LMS}{LMS}{Least-Mean-Squares}
\newacronym{NLMS}{NLMS}{Normalized Least-Mean-Squares}
\newacronym{AEC}{AEC}{Acoustic Echo Cancellation}
\newacronym{AES}{AES}{Acoustic Echo Suppression}
\newacronym{STFT}{STFT}{Short-Time Fourier Transform}
\newacronym{CTF}{CTF}{Convolutive Transfer Function}
\newacronym{FIR}{FIR}{Finite Impulse Response}
\newacronym{RLS}{RLS}{Recursive Least-Mean-Squares}
\newacronym{EIR}{EIR}{Echo-to-Interference Power Ratio}
\newacronym{VSS}{VSS}{Variable Step-Size}
\newacronym{MMSE}{MMSE}{Minimum Mean Square Error}
\newacronym{DNN}{DNN}{Deep Neural Network}
\newacronym{GRU}{GRU}{Gated Recurrent Unit}
\newacronym{MSE}{MSE}{Mean Square Error}
\newacronym{ERLE}{ERLE}{Echo Return Loss Enhancement}
\newacronym{AIR}{AIR}{Acoustic Impulse Response}
\newacronym{IR}{IR}{Impulse Response}
\newacronym{LEM}{LEM}{Loudspeaker-Enclosure-Microphone}
\newacronym{PESQ}{PESQ}{Perceptual Evaluation of Speech Quality}
\newacronym{DFT}{DFT}{Discrete Fourier Transform}
\newacronym{RTF}{RTF}{Real Time Factor}
\newacronym[firstplural=Power Spectral Densities (PSDs)]{PSD}{PSD}{Power Spectral Density}
\newacronym{BB}{BB}{Broadband}
\newacronym{NB}{NB}{Narrowband}
\newacronym{HB}{HB}{Hybrid}

\title{End-To-End Deep Learning-based Adaptation Control for Linear Acoustic Echo Cancellation}

\ifthenelse{\boolean{ieeeSubmission}}{
\author{Thomas Haubner,~\IEEEmembership{Student Member,~IEEE,} Andreas Brendel,~\IEEEmembership{Member,~IEEE,} \\and Walter Kellermann,~\IEEEmembership{Life Fellow,~IEEE}
\thanks{Manuscript received June x, 2023; revised Month~x, 2023; accepted Month~x, 2023. Date of publication Month~x, 2023; date of current version Month x, 2023. The associate editor coordinating the review of this manuscript and approving it for publication was Prof. X X. \\
Thomas Haubner and Walter Kellermann are with the chair of Multimedia Communications and Signal Processing (LMS), Friedrich-Alexander-Universität Erlangen-Nürnberg, D-91058 Erlangen, Germany. Andreas Brendel contributed to the work while he was at LMS. He is now with the Fraunhofer Institute for Integrated Circuits (IIS), D-91058 Erlangen, Germany. e-mail: \{thomas.haubner;andreas.brendel;walter.kellermann\}@fau.de \\
Digital Object Identifier}%
}%
}{
\author{Thomas Haubner,~\IEEEmembership{Student Member,~IEEE,} Andreas Brendel,~\IEEEmembership{Member,~IEEE,} \\and Walter Kellermann,~\IEEEmembership{Life Fellow,~IEEE}
\thanks{
This article has been submitted to \textit{IEEE/ACM Transactions on Audio, Speech, and Language Processing}. \\
Thomas Haubner and Walter Kellermann are with the chair of Multimedia Communications and Signal Processing (LMS), Friedrich-Alexander-Universität Erlangen-Nürnberg, D-91058 Erlangen, Germany. Andreas Brendel contributed to the work while he was at LMS. He is now with the Fraunhofer Institute for Integrated Circuits (IIS), D-91058 Erlangen, Germany. e-mail: \{thomas.haubner;andreas.brendel;walter.kellermann\}@fau.de \\
Digital Object Identifier
}%
}%
}

\ifthenelse{\boolean{ieeeSubmission}}{
\markboth{IEEE/ACM Transactions on Audio, Speech, and Language Processing,~Vol.~x, No.~x, June~2023}%
{Shell \MakeLowercase{\textit{et al.}}: A Sample Article Using IEEEtran.cls for IEEE Journals}
}{}


\maketitle

\begin{abstract}
The attenuation of acoustic loudspeaker echoes remains to be one of the open challenges to achieve pleasant full-duplex hands free speech communication. In many modern signal enhancement interfaces, this problem is addressed by a linear acoustic echo canceler which subtracts a loudspeaker echo estimate from the recorded microphone signal. To obtain precise echo estimates, the parameters of the echo canceler, i.e., the filter coefficients, need to be estimated quickly and precisely from the observed loudspeaker and microphone signals. For this a sophisticated adaptation control is required to deal with high-power double-talk and rapidly track time-varying acoustic environments which are often faced with portable devices. In this paper, we address this problem by end-to-end deep learning. In particular, we suggest to infer the step-size for a least mean squares frequency-domain adaptive filter update by a Deep Neural Network (DNN). Two different step-size inference approaches are investigated. On the one hand broadband approaches, which use a single DNN to jointly infer step-sizes for all frequency bands, and on the other hand narrowband methods, which exploit individual DNNs per frequency band. The discussion of benefits and disadvantages of both approaches leads to a novel hybrid approach which shows improved echo cancellation while requiring only small DNN architectures. Furthermore, we investigate the effect of different loss functions, signal feature vectors, and DNN output layer architectures on the echo cancellation performance from which we obtain valuable insights into the general design and functionality of DNN-based adaptation control algorithms. 
\end{abstract}


\section{Introduction}
The attenuation of loudspeaker echoes is a crucial component of any full-duplex hands-free speech communication interface \cite{haensler2004acoustic}. In general, there are two main algorithmic approaches to achieve this goal: \gls{AEC} and \gls{AES} \cite{haensler2004acoustic}. While echo cancelers attenuate the unwanted signal components by subtracting a loudspeaker echo estimate from the microphone signal, echo suppressors directly filter the microphone signal with a postfilter \cite{haensler2004acoustic}. In many modern speech enhancement algorithms, both approaches are combined by applying a postfilter to the echo-reduced error signal of the echo canceler \cite{haensler2004acoustic, enzner_acoustic_2014}. Despite recent progress by deep learning-based postfilters, the benefit of additionally using an echo canceler is significant \cite{aec_challenge_icassp_2021, aec_challenge_interspeech_2021, aec_challenge_icassp_2022}.
Yet, the full potential of an echo canceler is only obtained, if its model parameters are precisely inferred from the observed loudspeaker and microphone signals. For this, the continuous minimization of the microphone error signal power by gradient descent-based parameter updates has proven to be a powerful tool \cite{haykin_ad_filt_theory}. Yet, a sophisticated adaptation, i.e., step-size, control is required to rapidly track time-varying acoustic scenes and deal with high-power double-talk and noise \cite{MADER20001697}.
%

Adaptation control has evolved in the last decades from simple binary stall-or-adapt methods \cite{double_talk_detection_1,double_talk_detection_2}, which mainly considered double-talk, to advanced continuous step-size estimators which employ complex probabilistic models \cite{nitsch_frequency_selective_2000, ENZNER20061140, kuech_state_space_2014, benesty_nonparametric_2006, valin_adjusting_2007, nesta_batch_online, gunther_learning_2012, cheng_semi_blind_2021}. In particular, the inference of the model parameters by Kalman filtering \cite{ENZNER20061140, kuech_state_space_2014} or semi-supervised blind source separation \cite{nesta_batch_online, gunther_learning_2012, cheng_semi_blind_2021} can be considered as state of the art according to recent challenge results \cite{aec_challenge_icassp_2021, aec_challenge_interspeech_2021, aec_challenge_icassp_2022}. Yet, the performance of these traditional step-size estimators relies on the precise estimation of statistics of unobserved signals, e.g., the near-end speech and noise \glspl{PSD}, which are difficult to obtain in practice \cite{yang_improved_kf_with_fast_recovery}. Traditional estimators of these quantities often suffer from sensitive hyperparameter choices which do not generalize well to different acoustic scenes \cite{yang_improved_kf_with_fast_recovery}. To remedy this limitation, it has recently been proposed to use machine learning models to support traditional step-size estimators \cite{noise_dic_kf_haubner, syn_kal_dpf_haubner}. In particular, the approximation of the interference \gls{PSD} of the Kalman filter model, capturing near-end speech and noise, by non-negative dictionaries \cite{noise_dic_kf_haubner} and \glspl{DNN} \cite{syn_kal_dpf_haubner} has shown significant performance improvements relative to traditional, i.e., non-trainable, \gls{PSD} estimators. Besides the support of traditional step-size estimators by trainable \gls{PSD} models, machine learning has also been used to directly approximate optimum step-sizes of a time-domain \gls{NLMS} algorithm \cite{ivry_deep_nodate} or \gls{STFT}-domain recursive least squares algorithm \cite{rnn_step_size_ofer_schwartz}. Yet, despite significant performance improvements, it remains unclear whether machine learning-supported approximation of target step-sizes, e.g., a Kalman filter step-size with oracle statistics, is optimum w.r.t. echo cancellation performance. Furthermore, the loss function design for training the machine learning models is challenging, as the effect of \gls{PSD} or step-size estimation errors on the \gls{AEC} performance is complicated.
Therefore, it has recently been proposed to optimize \gls{DNN}-based step-size estimators directly w.r.t. the echo estimation quality which is typically termed end-to-end training \cite{dnn_fdaf_haubner, hybrid_deep_ad_aec_zhang, dnn_aec_bf_dpf, casebeer_meta-af_2023, wu_meta-learning_2022,  yang_low-complexity_2022, zhang_neuralkalman_2023}. The various approaches differ in the echo estimation model, feature design, \gls{DNN} integration and loss function design. While \cite{dnn_fdaf_haubner, dnn_aec_bf_dpf, casebeer_meta-af_2023, wu_meta-learning_2022} employ a time-domain \gls{FIR} echo estimation model which is optimized in the frequency domain, \cite{hybrid_deep_ad_aec_zhang,casebeer_meta-af_2023,yang_low-complexity_2022, zhang_neuralkalman_2023} consider a \gls{CTF} model (cf. e.g. \cite{multirate_systems_kellermann, diniz2012adaptive, avargel_system_2007}) which operates in the \gls{STFT} domain. Furthermore, while \cite{dnn_fdaf_haubner, dnn_aec_bf_dpf, casebeer_meta-af_2023, wu_meta-learning_2022} investigate linear echo estimation models, \cite{hybrid_deep_ad_aec_zhang,   yang_low-complexity_2022, zhang_neuralkalman_2023} consider additionally non-linear pre-processors to deal with hardware imperfections, e.g., loudspeaker distortions. Another key difference is given by the integration of the \gls{DNN} into the filter update. In \cite{casebeer_meta-af_2023,  wu_meta-learning_2022, yang_low-complexity_2022}, it is proposed to directly estimate the gradient, or parts of it, by complex-valued \glspl{DNN}. In contrast, \cite{dnn_fdaf_haubner, hybrid_deep_ad_aec_zhang, dnn_aec_bf_dpf, zhang_neuralkalman_2023} propose to infer only parameters of traditional step-size estimators while keeping the gradient direction of the respective filter update unchanged. Furthermore, various step-size inference architectures have been considered ranging from broadband approaches, which jointly estimate a frequency-selective step-size vector with a single \gls{DNN}, \cite{dnn_fdaf_haubner, hybrid_deep_ad_aec_zhang, dnn_aec_bf_dpf}, to narrowband approaches, which use individual \glspl{DNN} per frequency band \cite{casebeer_meta-af_2023, yang_low-complexity_2022, zhang_neuralkalman_2023}. In addition, joint control of frequency groups has been investigated \cite{wu_meta-learning_2022}. Finally, there are also echo cancellation algorithms which exploit no physical knowledge for designing the echo estimation model but instead use a completely data-driven \gls{DNN}-based estimator \cite{fingscheid_dnn_aec_1, fingscheid_dnn_aec_2,task_splitting_dnn}. 

In this paper, we provide a general view on deep learning-based adaptation control for linear \gls{AEC} and the differences between broadband and narrowband step-size inference architectures. By investigating their benefits and disadvantages, we introduce a new hybrid control strategy which combines the low \gls{DNN} parameter count of narrowband step-size estimators with the capability of broadband approaches to exploit information from the entire frequency spectrum. All three generic adaptation control approaches, i.e., narrowband, broadband, and hybrid, are compared for a variety of challenging \gls{AEC} scenarios. Furthermore, we analyze the effect of different features, loss functions and \gls{DNN} output layer designs on the echo cancellation performance. As a specific parameterization of our adaptation control algorithm is equivalent to a basic \gls{DNN}-based control of the \gls{NLMS} algorithm, the conclusions are broadly applicable. 
Additionally, to gain deeper insights into the \gls{DNN}-based step-size estimator, we analyze its internal states and step-size estimates by comparing them to the spectrograms of the individual microphone signal components. 
%
Finally, we provide the code for training the \glspl{DNN}\footnote{{\url{https://github.com/ThomasHaubner/e2e_dnn_ad_control_for_lin_aec}}} to simplify future comparison of different \gls{DNN}-based adaptation control algorithms and fuel the research on further developments.

The remainder of the paper is structured as follows: In Sec.~\ref{sec:lin_aec}, we define a signal model which describes a generic hands-free speech communication interface. Subsequently, the proposed deep learning-based adaptation control algorithms are introduced in Sec.~\ref{sec:ad_control} and their relation to traditional approaches is discussed. An experimental evaluation follows in Sec.~\ref{sec:exp} and conclusions are presented in Sec.~\ref{sec:conclusion}.

In the following, we use $f$ and $\tau$ as frequency and frame index of \gls{STFT}-domain quantities, respectively. The corresponding time-domain signals are denoted by an underline and $\kappa$ is used as the sample index. The real and imaginary components of a complex-valued number $z \in \mathbb{C}$ are denoted by $\mathcal{R}\{z\}$ and $\mathcal{I}\{z\}$, respectively, and its complex conjugate by $z^*$. The Euclidean norm of a vector is denoted by $\left|\left|\cdot \right| \right|$ and its transpose by $(\cdot)^{\text{T}}$. Additionally, we introduce the all-one vector $\boldsymbol{1}_F$ of dimension $F$ and the expectation operator $\mathbb{E}[\cdot]$. Finally, the $l$th diagonal element of a matrix is written as $\left[\cdot \right]_{ll}$.

\section{Linear Acoustic Echo Cancellation}
\label{sec:lin_aec}
This section introduces the signal model and the filter update which are commonly used for linear acoustic echo cancelers operating in the frequency domain as shown in Fig.~\ref{fig:lin_aec}.

\subsection{Signal Model}
\label{sec:sig_mod}
\begin{figure}[t]
	\centering
	\input{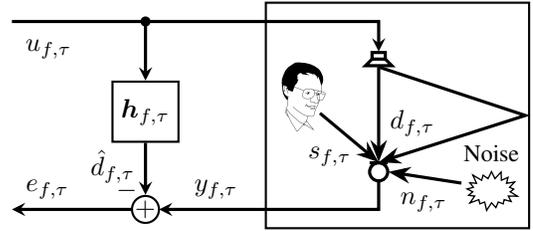}
	\caption{Block diagram of a linear acoustic echo canceler operating in the frequency domain.}
	\label{fig:lin_aec}
\end{figure}
We model the \ac{STFT}-domain microphone signal $y_{f,\tau}$ as a linear superposition of an echo component $d_{f,\tau}$, a near-end speech component $s_{f,\tau}$ and a background noise component $n_{f,\tau}$ as follows:
\begin{equation}
	y_{f,\tau} = d_{f,\tau} + s_{f,\tau} + n_{f,\tau} \in \mathbb{C}.
	\label{eq:mic_mod_base}
\end{equation}
The goal of \ac{AEC} is to obtain a precise echo estimate $\hat{d}_{f,\tau}$ which is subtracted from the microphone signal $y_{f,\tau}$ to obtain an ideally echo-free error signal
\begin{equation}
	e_{f,\tau} = 	y_{f,\tau}  - 	\hat{d}_{f,\tau}
 \label{eq:err_comp_prior}
\end{equation}
as shown in Fig.~\ref{fig:lin_aec}. In the following, we assume the popular \ac{CTF} echo estimation model \cite{multirate_systems_kellermann, diniz2012adaptive, avargel_system_2007} which suggests a linear \gls{FIR} filter model in each frequency band
\begin{equation}
	\hat{d}_{f,\tau} = {\boldsymbol{h}}_{f,\tau-1}^{\text{T}} \boldsymbol{u}_{f,\tau}
	\label{eq:ctf_echo_est}
\end{equation}
with the length-$L$ filter coefficient vector 
\begin{equation}
	\boldsymbol{h}_{f,\tau-1}^{\text{T}} = \begin{pmatrix}
		{h}_{0,f,\tau-1} & \dots & {h}_{L-1,f,\tau-1}
	\end{pmatrix} 
\end{equation}
and the \ac{STFT}-domain loudspeaker signal vector
\begin{equation}
	\boldsymbol{u}_{f,\tau}^{\text{T}} = \begin{pmatrix}
		{u}_{f,\tau} & \dots & {u}_{f,\tau-L+1}
	\end{pmatrix} .
\end{equation}

\subsection{Adaptive Filter Estimation}
\label{sec:af_est}
The filter coefficient vector $\boldsymbol{h}_{f,\tau-1}$ should ideally be updated such that the residual echo power $\mathbb{E}[| d_{f,\tau} - \hat{d}_{f,\tau} |^2 ]$ is minimized. Yet, as the echo statistics are not known in practice, filter coefficient updates are typically derived by iteratively minimizing the microphone error signal power\footnote{Note that the independent minimization of the frequency band-wise subband error powers requires sufficient suppression of cross-band aliasing to approximate the minimization of the time-domain error power.}
%
%
\begin{equation}
	{\psi}^{\text{EE}}_{f,\tau} = \mathbb{E}\left[\left| e_{f,\tau}\right|^2 \right] = \mathbb{E}\left[\left| d_{f,\tau} - \hat{d}_{f,\tau} + z_{f,\tau} \right|^2 \right]
			\label{eq:error_power}
\end{equation}
with $z_{f,\tau} = s_{f,\tau} + n_{f,\tau}$ denoting the interference component w.r.t. the filter estimation task. Note that the interference $z_{f,\tau}$ includes also all echo components which cannot be described by the linear \gls{CTF} model \eqref{eq:ctf_echo_est}, e.g., echoes resulting from late reverberation or loudspeaker non-linearities. The most popular filter optimization method is given by the \acrshort{LMS} update \cite{haykin_ad_filt_theory}
\begin{equation}
		{h}_{l,f,\tau} = {h}_{l,f,\tau-1} + {\mu}_{l,f,\tau} \left( {u}_{f,\tau-l}^* e_{f,\tau} \right)
		\label{eq:lms_update}
\end{equation}
with ${\mu}_{l,f,\tau} \geq 0$ being a tap-, frequency- and time-dependent step-size. Yet, by assuming the residual echo signal $d_{f,\tau} - \hat{d}_{f,\tau}$ and the interference component $z_{f,\tau}$ to be orthogonal, we easily conclude from \eqref{eq:error_power} that the error signal power $\mathbb{E}[|e_{f,\tau}|^2]$ is only a good approximation of the residual echo power $\mathbb{E}[| d_{f,\tau} - \hat{d}_{f,\tau} |^2 ]$ for time-frequency bins with a high \ac{EIR}
\begin{equation}
	\text{EIR}_{f,\tau} = 10 \log_{10} \frac{ 	\mathbb{E} \left[ \left| d_{f,\tau} \right|^2 \right]
	}{ 	\mathbb{E} \left[  \left| z_{f,\tau} \right|^2 \right] }.
\end{equation}
To remedy this limitation the step-size $\mu_{l,f,\tau}$ in the \acrshort{LMS} update \eqref{eq:lms_update} must be adapted to the respective signal statistics as will be discussed in the following.

\section{Adaptation Control}
\label{sec:ad_control}
In this section, we introduce and motivate the general concept of deep learning-based step-size estimation and discuss different generic variants.

\subsection{Traditional Approaches}
\label{sec:trad_ad_control}
We first introduce traditional model-based methods which will serve as a basis for the  deep learning-based approaches described in the sequel. The majority of these methods suggests a mapping of certain signal expectations, e.g., the loudspeaker and interference powers ${\psi}^{\text{{UU}}}_{f,\tau} = \mathbb{E}[||\boldsymbol{u}_{f,\tau}||^2]$ and ${\psi}^{\text{ZZ}}_{f,\tau} = \mathbb{E}[|{z}_{f,\tau}|^2]$, respectively, to the optimum step-size
\begin{equation}
	\mu_{l,f,\tau} \gets g^{\text{MB}}\left({\psi}^{\text{UU}}_{f,\tau}, {\psi}^{\text{ZZ}}_{f,\tau}, \dots  \right)
	\label{eq:fixed_step_size_mapping}
\end{equation}
%
%
with $g^{\text{MB}} \left(\cdot\right)$ denoting a deterministic function.
Among the most widely used traditional step-size selection strategies are 

a) the stall-or-adapt \ac{NLMS} step-size, e.g., \cite{double_talk_detection_1, double_talk_detection_2},
\begin{equation}
   \mu_{l,f,\tau} \gets \frac{m^{\text{DT}}_{\tau} }{ {\psi}^{\text{UU}}_{f,\tau}}\qquad \forall l \in \{0, \dots, L-1\}
	\label{eq:nlms_bin}
\end{equation}
with the binary parameter $m^{\text{DT}}_{\tau}$  being controlled by a double-talk detector, 

b) the error power-aware (EA) \ac{NLMS} step-size \cite{ia_nlms_greenberg,spriet2008feedback}
\begin{equation}
    \mu_{l,f,\tau} \gets \frac{m^{\text{EA-NLMS}} }{ {\psi}^{\text{UU}}_{f,\tau} +  {\psi}^{\text{EE}}_{f,\tau} }\qquad \forall l \in \{0, \dots, L-1\}
	\label{eq:nlms_ea}
\end{equation}
with the static hyperparameter $m^{\text{EA-NLMS}}>0$,

c) the minimum system-distance \ac{NLMS} step-size \cite{MADER20001697}
\begin{equation}
    \mu_{l,f,\tau} \gets \frac{ \mathbb{E} \left[\left| \left(\boldsymbol{h}_{f,\tau} -\boldsymbol{h}_{f,\tau}^{\text{MMSE}} \right)^{\text{T}}\boldsymbol{u}_{f,\tau}  \right|^2 \right]  }{ \mathbb{E} \left[ \left| \left( \boldsymbol{h}_{f,\tau} -\boldsymbol{h}_{f,\tau}^{\text{MMSE}} \right)^{\text{T}} \boldsymbol{u}_{f,\tau}\right|^2 \right]+ \psi^{\text{ZZ}}_{f,\tau}  } ~ \frac{1}{{\psi}^{\text{UU}}_{f,\tau}}, 
	\label{eq:nlms_vsss}
\end{equation}
for all $l\in \{0, \dots,L-1\}$ with the \ac{MMSE} filter estimate $\boldsymbol{h}_{f,\tau}^{\text{MMSE}}$, and 

d) the Kalman filter step-size \cite{ENZNER20061140, kuech_state_space_2014, LuisValero2019}
\begin{equation}
	\mu_{l,f,\tau} \gets \frac{{\Psi}^{\text{HH}}_{l,f,\tau}  }{ \sum_{l=0}^{L-1} {\Psi}^{\text{HH}}_{l,f,\tau} \left| {u}_{f,\tau-l}\right|^2 + \psi^{\text{ZZ}}_{f,\tau} }	
	\label{eq:kalman_nlms}
\end{equation}
with the filter estimation variances $ {\Psi}^{\text{HH}}_{0,f,\tau},\dots,{\Psi}^{\text{HH}}_{L-1,f,\tau}$. Note that except for the Kalman update \eqref{eq:kalman_nlms}, all considered adaptation control methods assume equal step-sizes for updating the filter coefficients within a time-frequency bin.
Furthermore, while \eqref{eq:nlms_bin} represents a binary broadband stall-or-adapt step-size, \eqref{eq:nlms_vsss} and \eqref{eq:kalman_nlms} yield continuous-valued frequency-selective step-sizes. All approaches have in common that they rely on a precise estimation of typically unknown expectations which often depend on non-observable signals, e.g., the interference power $\psi^{\text{ZZ}}_{f,\tau}$ or the \ac{MMSE} filter estimate $\boldsymbol{h}_{f,\tau}^{\text{MMSE}}$.
Furthermore, the claimed optimality of the respective step-sizes holds only under strong model assumptions, e.g., a linear Gaussian model for the Kalman filter update \eqref{eq:kalman_nlms}, whose validity is often questionable in practice. To remedy these limitations, we resort now to the concept of deep learning-based step-size estimation.

\subsection{Deep Learning-based Adaptation Control}
\label{sec:dnn_ad_control}
The general idea of deep learning-based step-size estimation is to replace the mapping $g^{\text{MB}}(\cdot)$ in \eqref{eq:fixed_step_size_mapping} by a \ac{DNN} \cite{dnn_fdaf_haubner, dnn_aec_bf_dpf, hybrid_deep_ad_aec_zhang}
\begin{equation}
	\boldsymbol{\mu}_{\tau} = \begin{pmatrix}
		\mu_{1,\tau} & \dots & \mu_{F,\tau}
	\end{pmatrix}^{\text{T}} \gets g^{\text{BB-DNN}}\left(\boldsymbol{\chi}_{\tau}, \boldsymbol{\zeta}_\tau; \boldsymbol{\theta}^{\text{BB}} \right)
	\label{eq:train_step_size_mapping}
\end{equation}
with parameter vector $\boldsymbol{\theta}^{\text{BB}}$ which infers the frame-dependent broadband step-size vector $\boldsymbol{\mu}_{\tau}$ from an observable feature vector $\boldsymbol{\chi}_{\tau}$ and internal \gls{DNN} state vector $\boldsymbol{\zeta}_{\tau}$. Note that we assume in Eq.~\eqref{eq:train_step_size_mapping} equal step-sizes within a time-frequency bin, i.e., $\forall l \in \{0, \dots, L-1\}: \mu_{f,\tau} = \mu_{f,l,\tau}$, which is common to many adaptation control approaches (cf.~Eqs.~\eqref{eq:nlms_bin} - \eqref{eq:nlms_vsss}).
%
\begin{figure*}[t]
	\centering

	\begin{subfigure}[b]{\textwidth}
		\centering
		\hspace*{-.0cm}\begin{tikzpicture}

	\begin{scope}[xshift=-.225cm, yshift=.25cm]
		\draw [very thick,fill=white] (0,-1) rectangle (4,1);
		\draw [step=.5, thick] (0,-1) grid (4,1);	
	\end{scope}
	
	\draw [very thick,fill=white] (0,-1) rectangle (4,1);
	\draw [step=.5, thick] (0,-1) grid (4,1);	
	\node [align=center] (featMap) at (-2.5, 0) {Spectro-Temporal \\ Feature Maps};

	\begin{scope}[xshift=-1.0cm, yshift=0.0cm]
		\draw [->, thick] (.55,.7) -- (.55,-.3) node [midway,left] {$f$};
		\draw [->, thick] (2.25, -1.25) -- (3.75, -1.25) node [midway,below] {$\tau$};
	\end{scope}
	
	\node (dnn1) at (8.5,.0) [draw, very thick,minimum width=1.20cm,minimum height=.5cm] {$g^{\text{BB-DNN}}(\cdot)$};

	\draw [very thick, ->, double] (dnn1.east) -- ($(dnn1)+(2.9,0)$) node [midway, above, pos=.45] {$\boldsymbol{\mu}_\tau$};

	%
	%

	\draw [ thick ,double, ->] (4,.75) -| ($(dnn1.north)+(-.0,.0)$) node [midway, above,pos=.25] {$\boldsymbol{\chi}_{1,\tau}$};
	\draw [ thick, double, ->] (4,.15) --  ($(dnn1.west)+(0.0,.15)$)  node [midway, above,pos=.62] {$\boldsymbol{\chi}_{2,\tau}$};
	\draw [ thick, double , ->] (4,-.75) -| ($(dnn1.south)+(-.0,-.0)$) node [midway, below,pos=.25] {$\boldsymbol{\chi}_{F,\tau}$};

	\begin{scope}[xshift=4.025cm, yshift=1.1cm]
		\draw [black,fill=black] (0,0) circle (.3mm);  
		\draw [black,fill=black] (-0.07,0.07) circle (.3mm);
		\draw [black,fill=black] (-0.14,0.14) circle (.3mm);	
	\end{scope}

	\node (dots1) at (6.2,-.2) {$\boldsymbol{\vdots}$};

\end{tikzpicture}
		\vspace*{-.4cm}
		\caption{Broadband}
		\label{fig:bb_dnn_vis}
	\end{subfigure}
	
	\vspace*{.2cm}
	
	\begin{subfigure}[b]{\textwidth}
		\centering
		\hspace*{-.0cm}\begin{tikzpicture}

	\begin{scope}[xshift=-.225cm, yshift=.25cm]
		\draw [very thick,fill=white] (0,-1) rectangle (4,1);
		\draw [step=.5, thick] (0,-1) grid (4,1);	
	\end{scope}
	
	\draw [very thick,fill=white] (0,-1) rectangle (4,1);
	\draw [step=.5, thick] (0,-1) grid (4,1);	
	\node [align=center] (featMap) at (-2.75, 0) {Spectro-Temporal \\ Feature Maps};

	\begin{scope}[xshift=-1.0cm, yshift=0.0cm]
		\draw [->, thick] (.55,.7) -- (.55,-.3) node [midway,left] {$f$};
		\draw [->, thick] (2.25, -1.25) -- (3.75, -1.25) node [midway,below] {$\tau$};
	\end{scope}
	
	\node (dnn1) at (8.5,.25) [draw, very thick,minimum width=1.20cm,minimum height=.5cm] {$g^{\text{NB-DNN}}(\cdot)$};
	
	\draw [ thick,double, ->] ($(dnn1)-(4.5,0)$) -- (dnn1.west) node [midway, above,pos=.5] {$\boldsymbol{\chi}_{f,\tau}$};
	\draw [ thick, ->] (dnn1.east) -- ($(dnn1)+(2.9,0)$) node [midway, above, pos=.5] {$\mu_{f,\tau}$};

	\node (dots1) at ($(dnn1)+(0,.80)$) {$\boldsymbol{\vdots}$};
	\node (dots2) at ($(dnn1)-(0,.75)$) {$\boldsymbol{\vdots}$};
	
	\begin{scope}[xshift=4.025cm, yshift=1.1cm]
		\draw [black,fill=black] (0,0) circle (.3mm);  
		\draw [black,fill=black] (-0.07,0.07) circle (.3mm);
		\draw [black,fill=black] (-0.14,0.14) circle (.3mm);	
	\end{scope}

	\node (dots1) at ($(dnn1)+(-1.8,.80)$) {$\boldsymbol{\vdots}$};
\node (dots2) at ($(dnn1)+(-1.8,-.75)$) {$\boldsymbol{\vdots}$};

	\node (dots1) at ($(dnn1)+(1.8,.80)$) {$\boldsymbol{\vdots}$};
\node (dots2) at ($(dnn1)+(1.8,-.80)$) {$\boldsymbol{\vdots}$};
	
\end{tikzpicture}
				\vspace*{-.4cm}
		\caption{Narrowband}
		\label{fig:nb_dnn_vis}
	\end{subfigure}
	
	\vspace*{.2cm}
	
	\begin{subfigure}[b]{\textwidth}
		\centering
		\hspace*{-.0cm}\begin{tikzpicture}

	\begin{scope}[xshift=-.225cm, yshift=.25cm]
		\draw [very thick,fill=white] (0,-1) rectangle (4,1);
		\draw [step=.5, thick] (0,-1) grid (4,1);	
	\end{scope}
	
	\draw [very thick,fill=white] (0,-1) rectangle (4,1);
	\draw [step=.5, thick] (0,-1) grid (4,1);	
	\node [align=center] (featMap) at (-2.5, 0) {Spectro-Temporal \\ Feature Maps};

	\begin{scope}[xshift=-1.0cm, yshift=0.0cm]
		\draw [->, thick] (.55,.7) -- (.55,-.3) node [midway,left] {$f$};
		\draw [->, thick] (2.25, -1.25) -- (3.75, -1.25) node [midway,below] {$\tau$};
	\end{scope}
	
	\node (dnn1) at (8.5,.25) [draw, very thick,minimum width=1.20cm,minimum height=.5cm] {$g^{\text{HB-DNN}}(\cdot)$};
	
	\draw [ thick, double, ->] ($(dnn1)+(-4.5,.15)$) -- ($(dnn1.west)+(0,.15)$) node [midway, above,pos=.75] {$\boldsymbol{\chi}_{f,\tau}$};
	\draw [ thick, ->] (dnn1.east) -- ($(dnn1)+(2.9,0)$) node [midway, above, pos=.5] {${\mu}_{f,\tau}$}; 
	
	%
	%

	\node (con) at (5.5,-.8) [draw, very thick,minimum width=1.20cm,minimum height=.5cm] {$g^{\text{BB-FEAT}}(\cdot)$};

	\draw [  thick, double, ->] (con.east) -- ($ (con.east) + (.2,0)$) |- ($(dnn1.west)+(0,-.1)$) node [pos=.22, right] {$\boldsymbol{\xi}_{\tau}$};
	
	
	\draw [ thick , double, dashed, ->] (4,.75) -| ($(con.north)+(.4,.0)$);
	\draw [ thick , double, dashed, ->] (4,.1) -| ($(con.north)+(-.0,.0)$);
	\draw [ thick , double, dashed,->] (4,-.2) -| ($(con.north)+(-.4,-.0)$);
	\draw [ thick , double, dashed, ->] (4,-.75) -- ($(con.west)+(-.0,+.05)$);
	
	\node (dots1) at ($(dnn1)+(0,.80)$) {$\boldsymbol{\vdots}$};
	\node (dots2) at ($(dnn1)-(0,.75)$) {$\boldsymbol{\vdots}$};
	
		\node (dots1) at ($(dnn1)+(1.8,.80)$) {$\boldsymbol{\vdots}$};
	\node (dots2) at ($(dnn1)+(1.8,-.80)$) {$\boldsymbol{\vdots}$};
	
	\begin{scope}[xshift=4.025cm, yshift=1.1cm]
		\draw [black,fill=black] (0,0) circle (.3mm);  
		\draw [black,fill=black] (-0.07,0.07) circle (.3mm);
		\draw [black,fill=black] (-0.14,0.14) circle (.3mm);	
	\end{scope}

	\node (dots1) at ($(dnn1)+(-1.2,.80)$) {$\boldsymbol{\vdots}$};
\node (dots2) at ($(dnn1)+(-1.2,-.75)$) {$\boldsymbol{\vdots}$};
	
\end{tikzpicture}
				\vspace*{-.4cm}
		\caption{Hybrid}
		\label{fig:hybrid_dnn_vis}
	\end{subfigure}
	\caption{Narrowband, broadband and hybrid \gls{DNN}-based adaptation control. While in broadband control a single \ac{DNN} is used to simultaneously infer a frequency-selective broadband step-size vector $\boldsymbol{\mu}_\tau$, narrowband and hybrid approaches use independent \glspl{DNN} to estimate scalar step-sizes $\mu_{f,\tau}$ per frequency band. While narrowband approaches exploit only information from the respective frequency band, hybrid methods use an additional low-dimensional feature vector $\boldsymbol{\xi}_{\tau}$ which is representative of the entire spectrum. The dashed arrows represent the dependency of the vector $\boldsymbol{\xi}_{\tau}$ on the observations of all frequency bands.}
\label{fig:nb_bb_hybrid_dnn_vis}
\end{figure*}
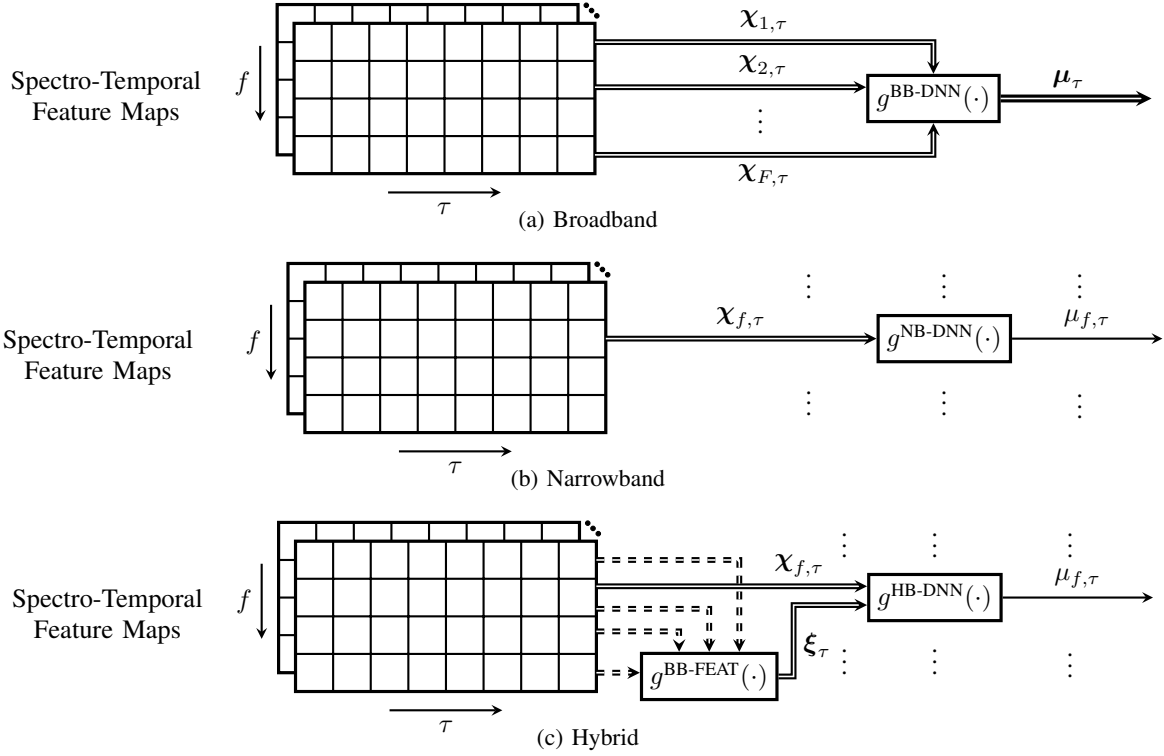

We now describe three different generic adaptation control strategies which differ in the structure of the \gls{DNN}-based mapping $g^{\text{BB-DNN}}(\cdot)$. For this we first introduce a decomposition of the broadband feature vector $\boldsymbol{\chi}_{\tau}$ into $F$ narrowband feature vectors $\boldsymbol{\chi}_{f,\tau}$ with $f=1,\dots,F$, which contain only information about the respective frequency bands as follows:
\begin{equation}
	\boldsymbol{\chi}_{\tau} = \begin{pmatrix}
		\boldsymbol{\chi}_{1,\tau}^{\text{T}} & \dots & \boldsymbol{\chi}_{F,\tau}^{\text{T}}
	\end{pmatrix}^{\text{T}}.
\end{equation}
This forms the input to the following three DNN-based step-size inference approaches:

a) {\em Broadband DNN.} The first and simplest method is to jointly infer all $F$ narrowband step-sizes $\mu_{1,\tau}, \dots,\mu_{F,\tau}$, contained in the broadband vector $\boldsymbol{\mu}_\tau$, by a single \gls{DNN} which exploits features from the entire frequency spectrum. This approach is straightforwardly related to the mapping \eqref{eq:train_step_size_mapping} and visualized in Fig.~\ref{fig:bb_dnn_vis}. 

b) {\em Narrowband DNN.}  In many traditional approaches (cf. Sec.~\ref{sec:trad_ad_control}) the narrowband step-size $\mu_{f,\tau}$ is computed only from statistics of the respective frequency band, i.e., without exploiting any inter-frequency dependencies. This motivates a decomposition of the general mapping $g^{\text{BB-DNN}}(\cdot)$ in \eqref{eq:train_step_size_mapping} into $F$ narrowband mappings \cite{casebeer_meta-af_2023, yang_low-complexity_2022, zhang_neuralkalman_2023}
\begin{equation}
		\mu_{f,\tau} \gets g^{\text{NB-DNN}}\left(\boldsymbol{\chi}_{f,\tau}, \boldsymbol{\zeta}_{f,\tau}; \boldsymbol{\theta}^{\text{NB}}  \right)
	\label{eq:train_step_size_mapping_nb}
\end{equation}
as shown in Fig.~\ref{fig:nb_dnn_vis}. Eq.~\eqref{eq:train_step_size_mapping_nb} suggests an inference of the narrowband step-size $\mu_{f,\tau}$ from a narrowband feature vector $\boldsymbol{\chi}_{f,\tau}$ with $\boldsymbol{\zeta}_{f,\tau}$ and ${\boldsymbol{\theta}}^{\text{NB}} $ denoting the internal state and parameter vectors of the \gls{DNN}, respectively. Note that we assume the \gls{DNN} parameter vector ${\boldsymbol{\theta}}^{\text{NB}} $ to be shared among different frequency bands which significantly reduces the number of trainable \gls{DNN} parameters. It furthermore offers the possibility to use a single trained \gls{DNN} for applications with different frequency resolutions which increases the flexibility. 

c) {\em Hybrid DNN.} A possible disadvantage of the narrowband relative to the broadband method is that it cannot exploit statistical inter-frequency dependencies for adaptation control. To remedy this limitation, while still benefiting from the low \gls{DNN} parameter count of narrowband approaches, we introduce a hybrid of both methods which is shown in Fig.~\ref{fig:hybrid_dnn_vis}. The key idea is to employ an individual mapping per frequency band, as suggested in the narrowband approach, yet, extend the narrowband feature vector $\boldsymbol{\chi}_{f,\tau}$ by a small number of additional hybrid features which are representative for the entire frequency spectrum. The respective hybrid step-size mapping is given by
\begin{equation}
	\mu_{f,\tau} \gets g^{\text{HB-DNN}}\left( \begin{pmatrix}
		\boldsymbol{\chi}_{f,\tau}^{\text{T}} & 	\boldsymbol{\xi}_{\tau}^{\text{T}}
	\end{pmatrix}^{\text{T}}, \boldsymbol{\zeta}_{f,\tau}; {\boldsymbol{\theta}}^{\text{HB}}  \right)
	\label{eq:train_step_size_mapping_hb}
\end{equation}
with the narrowband feature vector $\boldsymbol{\chi}_{f,\tau}$, the hybrid feature vector $\boldsymbol{\xi}_{\tau}$, the internal \gls{DNN} state vector $ \boldsymbol{\zeta}_{f,\tau}$ and the parameter vector $\boldsymbol{\theta}^{\text{HB}}$. In Fig.~\ref{fig:hybrid_dnn_vis} the computation of the hybrid feature vector $\boldsymbol{\xi}_{\tau}$ is indicated by the function $g^{\text{BB-FEAT}}(\cdot)$ which does not exhibit any trainable parameters. The benefit of the additional features critically depends on their design which is discussed in Sec.~\ref{sec:feat_vec}. 

Finally, we observe that all three approaches suggest a frequency-selective adaptation control scheme. An alternative is to enforce equivalent step-sizes for all frequency bands which is common to many time-domain \acrshort{LMS}-based filter updates. This approach can be directly incorporated into the broadband method \eqref{eq:train_step_size_mapping} by assuming
\begin{equation}
	\boldsymbol{\mu}_{\tau} \gets \mu_{\tau} \boldsymbol{1}_F
\end{equation} 
with the scalar step-size $\mu_{\tau}$ being provided by the \gls{DNN}.

\subsection{Feature Vectors}
\label{sec:feat_vec}
We will now discuss various choices for the narrowband and hybrid feature vectors $\boldsymbol{\chi}_{f,\tau}$ and $\boldsymbol{\xi}_{\tau}$, respectively, which are used to infer the step-sizes (cf. Eqs.~\eqref{eq:train_step_size_mapping} - \eqref{eq:train_step_size_mapping_hb}). The feature vectors can be computed from any subset of the following observable signals: The loudspeaker signal $u_{f,\tau}$, the microphone signal $y_{f,\tau}$, the error signal $e_{f,\tau}$ and the echo estimate $\hat{d}_{f,\tau}$. As we consider in this paper only real-valued \glspl{DNN}, each complex-valued signal must be transformed to the field of real numbers. A straightforward mapping is given by stacking the real and imaginary parts of a subset of the complex-valued signals, e.g., the loudspeaker and microphone signals $u_{f,\tau}$ and $y_{f,\tau}$, respectively,
\begin{equation}
	\boldsymbol{\chi}_{f,\tau}^{\text{T}} \gets
	\begin{pmatrix}
		\mathcal{R} \{ u_{f,\tau} \} &  \mathcal{I} \{ u_{f,\tau} \}  &
		\mathcal{R} \{ y_{f,\tau} \} &  \mathcal{I} \{ y_{f,\tau} \} 
	\end{pmatrix}.
\label{eq:re_im_feat}
\end{equation}
Many traditional adaptation control approaches (cf. Sec.~\ref{sec:trad_ad_control}) mainly rely on the signal magnitude, i.e., discard the phase information. This motivates a non-linear mapping by the magnitude operator and an additional logarithmic transformation with exemplary narrowband feature vectors being given by
\begin{align}
	\boldsymbol{\chi}_{f,\tau}^{\text{T}} &\gets 
	\begin{pmatrix}
		\left|  u_{f,\tau} \right| &  	\left|  y_{f,\tau} \right|
	\end{pmatrix},
	\label{eq:mag_feat} \\
	\boldsymbol{\chi}_{f,\tau}^{\text{T}}& \gets 
	\begin{pmatrix}
		\log_{10} \left( 	\left|  u_{f,\tau} \right|  + \delta^\text{FEAT} \right) &  		\log_{10} \left( 	\left|  y_{f,\tau} \right|  + \delta^\text{FEAT} \right)
	\end{pmatrix},
\label{eq:log_abs_feat_trans}
\end{align}
respectively, with $\delta^\text{FEAT}>0$ being a small constant to avoid numerical instabilities. Note that in principle any combination of signals and transformations can be used to define a narrowband feature vector $\boldsymbol{\chi}_{f,\tau}$. Yet, some combinations might be less promising, considering for example the linear relationship of the microphone, error and echo estimates. 

We will now discuss the design of the hybrid feature vector $\boldsymbol{\xi}_{\tau}$ which should be representative for the entire frequency spectrum at time frame $\tau$. Similar to the narrowband feature vector $\boldsymbol{\chi}_{f,\tau}$, it can be computed from a subset of the complex-valued signals $u_{f,\tau}$, $y_{f,\tau}$, $e_{f,\tau}$ and $\hat{d}_{f,\tau}$, yet, now considering all frequency bands. To condense the respective broadband vectors to low-dimensional real-valued scalars, we consider the arithmetic magnitude averages \cite{schwarz_spectral_2013}
\begin{equation}
	\bar{y}_\tau = \frac{1}{F} \sum_{f=1}^{F} \left| y_{f,\tau} \right|
	\label{eq:hybrid_features}
\end{equation}
with $\bar{e}_\tau$ and $\bar{\hat{d}}_\tau$ being defined accordingly. By stacking a subset of the magnitude averages $\bar{y}_\tau$, $\bar{e}_\tau$ and $\bar{\hat{d}}_\tau$ into a vector, we obtain an exemplary hybrid feature vector $\boldsymbol{\xi}_{\tau}$. As the feature \eqref{eq:hybrid_features} is independent of the frequency resolution, the hybrid method keeps the flexibility benefit, i.e., application to different frequency resolutions, of the narrowband approach. Finally, we note that all feature vectors are normalized to zero mean and unit variance with the respective statistics being estimated during training.

\subsection{Output Layer Design of the \glspl{DNN}}
\label{sec:out_layer_design}
\begin{figure}[tb]
	\centering
		\begin{tikzpicture}[node distance=2.0cm, >=stealth]

	\node (featSig) at (0,0) {};
	\node [rectangle, draw, thick, right of=featSig, minimum height=0.7cm] (ffGru) {{DNN}};

	\node [rectangle, draw, thick, right of=ffGru, yshift=-.0cm, xshift=1.5cm] (ffOutMu) {{\large$\frac{m^{\text{DNN-$\mu$}}_{f,\tau} }{ \hat{\psi}^{\text{UU}}_{f,\tau} +  \left|  m^{\text{DNN-e}}_{f,\tau} e_{f,\tau} \right|^2 }$}};
	
	\node [right of=ffOutMu,xshift=1cm] (out) {};
	
	\draw [thick, ->] (featSig) -- (ffGru.west) node [midway, above] {${\boldsymbol{\chi}}_{f,\tau}$};
	
	\draw [thick, ->] (ffOutMu.east) -- (out) node [midway, above] {${\mu}_{f,\tau}$};

		\draw [thick, ->] ($(ffGru.east)+(0,.15)$) -- ($(ffOutMu.west)+(0,.15)$) node [midway, above] {${m}_{f,\tau}^{\text{DNN-}\mu}$};
		\draw [thick, ->] ($(ffGru.east)-(0,.15)$) -- ($(ffOutMu.west)-(0,.15)$) node [midway, below] {${m}_{f,\tau}^{\text{DNN-e}}$};
	

	
		\draw [thick, ->] ($(ffGru.north)+(.3,0)$) to [out=75,in=0] ($(ffGru.north)+(.0,.4)$) to [out=180,in=105] ($(ffGru.north)+(-.3,0)$) ;
	\node (delay) at ($(ffGru.north)+(.0,.65)$)  {$\boldsymbol{\zeta}_{f,\tau}$};
	
\end{tikzpicture}
	\caption{Block diagram of the proposed \ac{DNN}-controlled narrowband step-size estimation approach with the \gls{DNN}-provided masks ${m}_{f,\tau}^{\text{DNN-}\mu}$ and ${m}_{f,\tau}^{\text{DNN-e}}$, the feature vector ${\boldsymbol{\chi}}_{f,\tau}$ and the internal \gls{DNN} state vector $\boldsymbol{\zeta}_{f,\tau}$.}
	\label{fig:vis_overal_fig}
    \vspace*{-.2cm}
\end{figure}
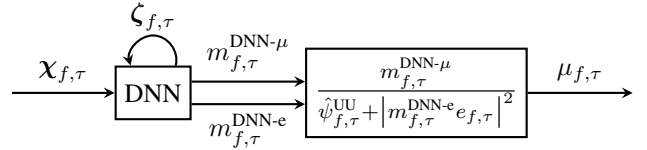

In Sec.~\ref{sec:dnn_ad_control}, we discussed the key idea of estimating the step-size ${\mu}_{f,\tau}$, which controls the \gls{CTF} filter adaptation (cf. Eq.~\eqref{eq:lms_update}), by a \gls{DNN}. Yet, due to the non-whiteness and non-stationarity that many acoustic signals exhibit \cite{vary2006digital}, a direct estimation of the \acrshort{LMS} step-size $ \mu_{f,\tau}$ requires the \gls{DNN} to cover a large numerical range in its last layer. To remedy this limitation we enforce a specific structure in the output layer of the \gls{DNN}. To motivate this structure we first consider the optimum \ac{NLMS} step-size \eqref{eq:nlms_vsss}. We observe that the second part of the product, i.e., the loudspeaker power normalization, is easily estimated by recursive averaging
\begin{equation}
	\hat{\psi}_{f,\tau}^{\text{UU}} = \lambda_U \hat{\psi}_{f,\tau-1}^{\text{UU}} + (1-\lambda_U)  \left|\left| \boldsymbol{u}_{f,\tau}\right|\right|^2
	\label{eq:ls_pow_est}
\end{equation}
whereas the first part is difficult to estimate, yet, limited to the range $[0,1]$. Thus, to address the numerical range difficulty, a straightforward idea is to use a sigmoid activation in the output layer of the \gls{DNN} and subsequently normalize the outcome by the loudspeaker power estimate \eqref{eq:ls_pow_est} \cite{dnn_fdaf_haubner, hybrid_deep_ad_aec_zhang}. Yet, as the basic \ac{NLMS} update is not interference-robust, the \ac{DNN} training might be complicated as no guidance is available. In contrast, the traditional step-size estimators \eqref{eq:nlms_ea} - \eqref{eq:kalman_nlms} have already been derived with the goal of interference-robustness. To exploit this additional domain knowledge, and therefore potentially simplifying the \ac{DNN} training, we enforce the following output layer, and thus step-size, structure \cite{dnn_fdaf_haubner}
\begin{equation}
	\mu_{f,\tau} \gets \frac{m^{\text{DNN-$\mu$}}_{f,\tau} }{ \hat{\psi}^{\text{UU}}_{f,\tau} +  \left|  m^{\text{DNN-e}}_{f,\tau} e_{f,\tau} \right|^2+ \delta^{\text{VSS}} }
	\label{eq:dnn_ea_nlms}
\end{equation}
with the \ac{DNN}-provided masks $m^{\text{DNN-$\mu$}}_{f,\tau}$ and $m^{\text{DNN-e}}_{f,\tau}$ being limited to the range from $0$ to $1$ and the small regularization constant $\delta^{\text{VSS}}>0$. Note that the output layer structure \eqref{eq:dnn_ea_nlms} is closely related to the error power-aware \ac{NLMS} step-size \eqref{eq:nlms_ea} with the \ac{DNN} providing a frequency- and frame-selective hyperparameter $m^{\text{EA-NLMS}}$ and controlling the error power normalization by $m^{\text{DNN-e}}_{f,\tau}$. 
By choosing $m^{\text{DNN-e}}_{f,\tau}$ $=$ $0$ we obtain a \gls{DNN}-controlled \gls{NLMS} update similar to \cite{hybrid_deep_ad_aec_zhang}. The overall approach, including a narrowband \gls{DNN}, is shown in Fig.~\ref{fig:vis_overal_fig}.

\subsection{Loss Functions}
\label{sec:loss_func}
\begin{figure}[htbp]
	\centering
		
\begin{tikzpicture}[node distance=1.49cm, >=stealth]
	
	\def\ySpaceAA{.2}
	\def\ySpaceBB{-.2}
	\def\ySpaceAAC{-1.9}

	\def\innerSepAA{.05}
	\def\innerSepB{.05}
	\def\innerSepC{.1}
	\def\innerSepD{.175}

	\node [circle, draw,inner sep=\innerSepB cm, thick] (hatW0) at (0,0) {$\textcolor{black}{{{\boldsymbol{h}}}_{f,0}}$};
	
	\node [circle, draw, inner sep=\innerSepB cm, right of=hatW0, thick] (hatW1)  {$\textcolor{black}{{{\boldsymbol{h}}}_{f,1}}$};
	\node [circle, draw, inner sep=\innerSepB cm, right of=hatW1, thick] (hatW2) {$\textcolor{black}{{{\boldsymbol{h}}}_{f,2}}$};
	\node [right of=hatW2] (hatW3) {\large$\dots$};
	\node [circle, draw, inner sep=\innerSepB cm, right of=hatW3, thick] (hatW4) {$\textcolor{black}{{{\boldsymbol{h}}}_{f,T}}$};
	
	\node [circle, draw,inner sep=\innerSepAA cm, right of=hatW0, above of=hatW0, yshift=\ySpaceBB cm, thick] (lambda1) {\textcolor{black}{${\mu}_{f,1}$}};
	\node [circle, draw,inner sep=\innerSepAA cm, right of=hatW1, above of=hatW1, yshift=\ySpaceBB cm, thick] (lambda2) {\textcolor{black}{${\mu}_{f,2}$}};
	\node [right of=lambda2,yshift=0 cm] (lambda3) {\large$\dots$};
	\node [circle, draw,inner sep=\innerSepAA cm, right of=lambda3, yshift=-.1 cm, thick] (lambda4) {\textcolor{black}{${\mu}_{f,T}$}};
	
	\node [circle, draw,inner sep=.3 cm, right of=hatW0, below of=hatW0, yshift=\ySpaceAA cm, thick] (ups1) {};
	\node [circle, draw,inner sep=.3 cm, right of=hatW1, below of=hatW1, yshift= \ySpaceAA cm, thick] (ups2) {};
	\node [right of=ups2, yshift=.0cm] (ups3) {\large$\dots$};
	\node [circle, draw,inner sep=.3 cm, right of=ups3, yshift=.0cm, thick] (ups4) {};

   \node (tdEcho1Text) at (ups1) {\textcolor{black}{${{\hat{d}}}_{f,1}$}};
   \node (tdEcho2Text) at (ups2) {\textcolor{black}{${{\hat{d}}}_{f,2}$}};
   \node (tdEcho3Text) at (ups4) {\textcolor{black}{${{\hat{d}}}_{f,T}$}};
 
	
	\node [circle, dashed, draw,inner sep=0.10cm,  above of=lambda1, xshift=2.25cm, yshift=-.4cm, thick] (theta) {\textcolor{black}{$\boldsymbol{\theta}$}};
	
	\node [rectangle, draw,inner sep=0.10cm,  below of=ups4, xshift=-2.25cm, yshift=.43cm, very thick, minimum width=6cm, minimum height=.6cm] (istft) {Inverse STFT};

	\node [circle, draw,inner sep=.3cm, right of=hatW0, below of=hatW0, yshift=\ySpaceAAC cm, thick] (tdEcho1) {};
	\node [circle, draw,inner sep=.3 cm, right of=hatW1, below of=hatW1, yshift= \ySpaceAAC cm, thick] (tdEcho2) {};
	\node [right of=tdEcho2, yshift=.0cm] (tdEcho5) {\large$\dots$};
	\node [circle, draw,inner sep=.3 cm, right of=hatW2, below of=hatW3, yshift= \ySpaceAAC cm, thick] (tdEcho3) {};

   \node (tdEcho1Text) at (tdEcho1) {\textcolor{black}{${{\hat{\underline{d}}}}_{1}$}};
   \node (tdEcho2Text) at (tdEcho2) {\textcolor{black}{${{\hat{\underline{d}}}}_{2}$}};
   \node (tdEcho3Text) at (tdEcho3) {\textcolor{black}{${{\hat{\underline{d}}}}_{K}$}};
 

	\draw [thick, ->] (hatW0) -- (hatW1);
	\draw [thick, ->] (hatW1) -- (hatW2);
	\draw [thick, ->] (hatW2) -- (hatW3);
	\draw [thick, ->] (hatW3) -- (hatW4);
	\draw [thick, ->] (lambda1) -- (hatW1);
	\draw [thick, ->] (lambda2) -- (hatW2);
	\draw [thick, ->] (lambda4) -- (hatW4);
	
	
	\draw [thick, ->] (ups1.south) -- ($(ups1.south)-(0,.3)$);
	\draw [thick, ->] (ups2.south) -- ($(ups2.south)-(0,.3)$);
	\draw [thick, ->](ups4.south) -- ($(ups4.south)-(0,.3)$);
	
	\draw [thick, ->] ($(tdEcho1.north)+(0,0.3)$) -- (tdEcho1.north);
	\draw [thick, ->]  ($(tdEcho2.north)+(0,0.3)$) -- (tdEcho2.north);
	\draw [thick, ->] ($(tdEcho3.north)+(0,0.3)$) -- (tdEcho3.north);

	\draw [thick, ->] (hatW0) -- (ups1);
	\draw [thick, ->] (hatW1) -- (ups2);
	\draw [thick, ->] (hatW3) -- (ups4);
	
	\draw [thick, ->] (hatW0) -- (lambda1);
	\draw [thick, ->] (hatW1) -- (lambda2);
	\draw [thick, ->] (hatW2) -- (lambda3);
	\draw [thick, ->] (hatW3) -- (lambda4);
	
	\draw [thick, dashed, ->] (theta) -- (lambda1);
	\draw [thick, dashed, ->] (theta) -- (lambda2);
	\draw [thick, dashed, ->] (theta) -- (lambda3);
	\draw [thick, dashed, ->] (theta) -- (lambda4);
	
\end{tikzpicture}
	\caption{Relation of the \gls{DNN} parameter vector $\boldsymbol{\theta}$ to the \gls{CTF} filter estimates $\boldsymbol{h}_{f,\tau}$, \ac{STFT}-domain echo estimates $\hat{d}_{f,\tau}$ and time-domain echo estimates $\hat{\underline{d}}_{\kappa}$.} 
	\label{fig:rel_dnn_par_vec_loss}
    \vspace*{-.1cm}
\end{figure}
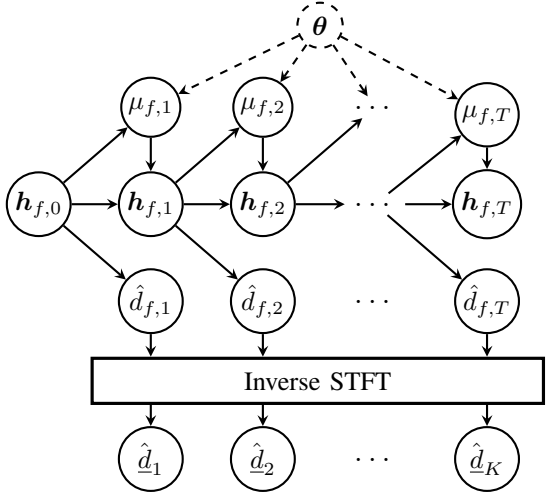

We now discuss different loss functions for optimizing the broadband, narrowband and hybrid \ac{DNN} parameter vectors $\boldsymbol{\theta}^{\text{BB}} $, ${\boldsymbol{\theta}}^{\text{NB}}$, and ${\boldsymbol{\theta}}^{\text{HB}}$, respectively. Note that we use $\boldsymbol{\theta}$ as representative for all three parameter vectors in this section. The relation of the \gls{DNN} parameter vector $\boldsymbol{\theta}$ to the \gls{CTF} filter estimates $\boldsymbol{h}_{f,\tau}$, \ac{STFT}-domain echo estimates $\hat{d}_{f,\tau}$ and time-domain echo estimates $\hat{\underline{d}}_{\kappa}$ with the frequency index $f\in \{1,\dots,F\}$, the frame index $\tau\in \{1,\dots,T\}$ and the time-domain sample index $\kappa\in \{1,\dots,K\}$, respectively, is shown in Fig.~\ref{fig:rel_dnn_par_vec_loss}.
In general, there are two types of loss functions which differ in their optimization targets: 1) Approximation of oracle target step-sizes, 2) Approximation of the desired echo signal.
%
%
Possible choices for the desired target step-sizes could be the minimum system distance \ac{NLMS} or Kalman step-sizes given in Eqs.~\eqref{eq:nlms_vsss} and \eqref{eq:kalman_nlms}. Yet, this approach is challenged by the difficulty to design a proper distance measure as the effect of estimation errors on the echo cancellation performance is complicated. Furthermore, as optimum model-based step-sizes are typically derived by certain signal assumptions, e.g., a linear Gaussian observation model within the Kalman filter framework, their \textit{optimality} cannot be guaranteed for realistic data. Therefore, we resort in this paper to the second approach which represents an end-to-end optimization strategy \cite{dnn_fdaf_haubner, hybrid_deep_ad_aec_zhang, dnn_aec_bf_dpf, casebeer_meta-af_2023, wu_meta-learning_2022,  yang_low-complexity_2022, zhang_neuralkalman_2023} as it directly quantifies the effect of the step-size estimate on the echo cancellation performance. Considering the different signal domains, i.e., time or frequency domain, we always investigate pairs of loss functions in the following. A first naive choice is given by the \ac{MSE} losses 
\begin{align}
	\mathcal{J}^{\text{FD-MSE}}(\boldsymbol{\theta}) & = \frac{1}{TF} \sum_{\tau,f=1}^{T,F} \left| {d}_{f,\tau} - \hat{{d}}_{f,\tau} \right|^2 \label{eq:fd_mse_loss} \\
	\mathcal{J}^{\text{TD-MSE}}(\boldsymbol{\theta}) & = \frac{1}{K} \sum_{\kappa=1}^{K} \left| \underline{d}_\kappa - \hat{\underline{d}}_\kappa \right|^2 \label{eq:td_mse_loss}
\end{align} 
with $\underline{d}_\kappa$ and $\hat{\underline{d}}_\kappa$ denoting the time-domain echo and estimated echo signals at sample index $\kappa$, respectively, and $K,T$ denoting the number of time-domain samples and frames, respectively. It should be noted that while the frequency-domain loss \eqref{eq:fd_mse_loss} suggests an independent residual echo power minimization per frequency band, its time-domain counterpart \eqref{eq:td_mse_loss} couples the frequency-wise echo estimates by the inverse \gls{STFT}. 
In addition, we examine the negated logarithmic \ac{ERLE}-type losses
\begin{align}
			\mathcal{J}^{\text{FD-ERLE}}(\boldsymbol{\theta}) & = - \log_{10} \frac{ \delta^\text{LOSS} + \frac{1}{TF} \sum_{\tau,f=1}^{T,F} \left| {d}_{f,\tau} \right|^2}{ \delta^\text{LOSS} +\frac{1}{TF}  \sum_{\tau,f=1}^{T,F} \left| {d}_{f,\tau} - \hat{d}_{f,\tau } \right|^2}
	\label{eq:fd_erle_loss} \\
		\mathcal{J}^{\text{TD-ERLE}}(\boldsymbol{\theta}) & = - \log_{10} \frac{\delta^\text{LOSS} + \frac{1}{K}  \sum_{\kappa=1}^{K} \left| \underline{d}_\kappa \right|^2}{\delta^\text{LOSS} + \frac{1}{K}  \sum_{\kappa=1}^{K}  \left| \underline{d}_\kappa - \hat{\underline{d}}_\kappa \right|^2}
		\label{eq:td_erle_loss}
\end{align}
with $\delta^\text{LOSS} >0$ being a regularization factor to avoid numerical instabilities. 
Note that the echo powers in the numerators of \eqref{eq:fd_erle_loss} and \eqref{eq:td_erle_loss} do not depend on the \ac{DNN} parameter vector $\boldsymbol{\theta}$ and could therefore be discarded. Nevertheless we keep it here for its analogy with the commonly-used \gls{ERLE}. Note that the respective \ac{MSE} and \ac{ERLE} losses essentially differ in the additional transformation by a logarithm.

\section{Experiments}
\label{sec:exp}
We will now compare the various adaptation control approaches (cf. Sec.~\ref{sec:dnn_ad_control}), signal feature vector choices (cf. Sec.~\ref{sec:feat_vec}), output layer designs (cf. Sec.~\ref{sec:out_layer_design}) and loss functions (cf. Sec.~\ref{sec:loss_func}) w.r.t. each other and relative to various well-known traditional adaptation control algorithms (cf.~\ref{sec:trad_ad_control}) for a variety of challenging acoustic scenarios.

\subsection{Experimental Setup}
\label{sec:exp_set}
In this section, we describe the experimental setup which was used to train the \glspl{DNN} and compare the different adaptation control algorithms. As introduced in Sec.~\ref{sec:sig_mod}, we consider the microphone signal to be composed of a loudspeaker echo, near-end speech and background noise component (cf. Eq.~\eqref{eq:mic_mod_base}). The near-end speech signals for the training and test sets were sampled randomly from the \textit{real} ($12$~h) and \textit{blind test} ($1.2$~h) data sets, respectively, of the \textit{Microsoft AEC Challenges} \cite{aec_challenge_icassp_2021, aec_challenge_interspeech_2021, aec_challenge_icassp_2022}, which contains recordings from a variety of acoustic environments and devices. The background noise signals were sampled randomly from the training ($150$~h) and test sets ($17$~h) of the \textit{DNS} \cite{dubey2023icasspDNSchallenge} and the \textit{CHiME3} \cite{chime_3} challenges, respectively. The echo signals were computed by a linear convolution of randomly-selected \gls{LEM} \glspl{IR} with randomly-selected music or speech loudspeaker signals. The \gls{LEM} \glspl{IR} for computing the echo signals in the training data were sampled randomly from the \textit{Microsoft AEC Challenges} {\cite{aec_challenge_icassp_2021, aec_challenge_interspeech_2021, aec_challenge_icassp_2022}} and \textit{MIT Survey} \cite{mit_ir_survey} data sets and comprised a total of {$1134$} measured \glspl{IR} from a variety of acoustic scenes and devices. Note that \cite{aec_challenge_icassp_2021, aec_challenge_interspeech_2021, aec_challenge_icassp_2022} provide only sweep measurements and the respective \glspl{IR} must be computed in a pre-processing step. The \gls{LEM} \glspl{IR} for generating the test echo data were taken from a disjoint subset of \cite{aec_challenge_icassp_2021, aec_challenge_interspeech_2021, aec_challenge_icassp_2022} and represented in total {$300$} different acoustic scenes and devices. The loudspeaker speech and music signals for training were taken from the training subsets of \cite{libri_speech} {($11.5$~h and $919$ speakers)} and \cite{gtzan, gtzan_sturm} ({$5.5$}~h from different genres), respectively. The test speech signals were taken from \cite{libri_speech, librivox_website} ($2.6$~h and {$209$} speakers) and the music test signals were taken from the test subset of \cite{gtzan, gtzan_sturm} ({$2.4$ h}). Note that we refrained from directly taking the recorded echo signals from \cite{aec_challenge_icassp_2021, aec_challenge_interspeech_2021, aec_challenge_icassp_2022} as only speech far-end signals are available and many recordings contain high-level fan noise which cannot be attenuated by an echo canceler and thus complicate a proper evaluation. The respective fan noise is typically dealt with by a post filter \cite{haensler2004acoustic} which we do not consider in this paper, yet, could easily be appended to the discussed echo cancelers\footnote{Note that fan noise is still considered in the background noise signals from {\cite{dubey2023icasspDNSchallenge}} and thus its effect on the adaptation control performance can be assessed.}. To investigate the tracking behaviour of the algorithms, which is of particular interest for mobile devices, we also considered time-varying acoustic scenes. For this, we switched the \gls{LEM} \glspl{IR} in $90$\% of the signals after randomly-selected scene durations in the range from $3$~s to $6$~s. To include both, hard and smooth transitions, a linear fading with a randomly-selected duration between $0$~s to $1$~s has been considered. Furthermore, two thirds of the loudspeaker and near-end speech signals were pre-processed by a temporal masking with random onsets and offsets to mimic conversations with non- and partially overlapping activity. Finally, the signals were scaled to obtain random \textit{near-end-to-echo} and \textit{near-end-to-echo-and-noise} power ratios in the ranges $[-10, 10]$ dB and $[20, 40]$~dB, respectively, and unit loudspeaker/microphone variances. Note that the respective power estimates were computed only over segments with considerable signal activity to limit the effect of pauses. All signals were sampled at a sampling frequency of $16$~kHz and limited to a duration of $8$~s. In total, we used $6$~h of training data, $1.5$~h of validation data and $1$~h of testing data. Note that the \gls{LEM} \glspl{IR}, speech, music and noise signals of the test data are disjoint from the training/validation data sets.

\subsection{Algorithmic Settings}
\label{sec:dnn_architecture}
We now discuss the parameter settings of the various algorithms. The frame shift and DFT length of the \gls{STFT} were chosen to $128$ and $512$, respectively, and the Hamming window has been applied prior to transformation\footnote{Note that a quarter-frameshift is chosen to attain sufficient cross-band attenuation.}.
The \gls{CTF} filter length $L$ was set to $8$ which covers approximately $88$ ms of the corresponding \gls{LEM} \glspl{IR}. 

\subsubsection{Deep Learning-based Adaptation Control}
The \gls{DNN} architecture was composed of a fully connected layer with a leaky RELU activation, two stacked \gls{GRU} layers and two parallel fully connected layers with sigmoid activations which map from the \gls{GRU} states to the different adaptation control masks $m^{\text{DNN-$\mu$}}_{f,\tau}$ and $m^{\text{DNN-e}}_{f,\tau}$, respectively that define the step-size (cf. Eq.~\eqref{eq:dnn_ea_nlms}). This commonly used architecture was chosen to encode important information about the near-end speech/background noise activity and filter convergence state from the feature sequence within the \gls{GRU} states. 
Note that the proposed method is not limited to this architecture and other architectures can straightforwardly be embedded. The recursive loudspeaker power estimation factor and the various regularization constants are chosen to $\lambda_U=0.9$, $\delta^\text{FEAT}=\delta^\text{LOSS}= 10^{-12}$ and $\delta^{\text{VSS}} =10^{-3}$, respectively. The dimension of the \gls{GRU} states was set to {$128$} for the broadband step-size estimator (cf. Eq.~\eqref{eq:train_step_size_mapping}) and to $64$ for the narrowband and hybrid estimators (cf. Eqs.~\eqref{eq:train_step_size_mapping_nb} and \eqref{eq:train_step_size_mapping_hb}), respectively. Note that the \gls{GRU} dimensions have been chosen differently to account for the respective task difficulties, i.e., estimation of a step-size vector $\boldsymbol{\mu}_{\tau}$ in comparison to a scalar step-size $\mu_{f,\tau}$. The \glspl{DNN} were trained for $60$ epochs with the ADAM optimizer \cite{kingma2014adam}, a step-size of $10^{-3}$ and an early stopping after $20$ epochs with no performance improvement on the validation data. Furthermore, the ADAM step-size was decreased by one-half after five consecutive epochs with no performance improvement, and Euclidean norm-based gradient clipping with a threshold of $0.5$ was applied. The batch size was chosen as $4$ and $32$ for the narrowband/hybrid and broadband methods, respectively, to account for hardware memory requirements. Finally, the optimum model was selected based on the lowest training loss on the validation data.

\subsubsection{Traditional and Oracle Adaptation Control}
To assess the performance of the proposed deep learning-based adaptation control algorithms, we consider various widely-used traditional step-size estimation algorithms. In particular, we consider the error power-aware \gls{NLMS} algorithm \cite{ia_nlms_greenberg,spriet2008feedback}, denoted by \makebox{\textit{EA-NLMS}} (cf. Eq.~\eqref{eq:nlms_ea}), and the narrowband \gls{CTF} Kalman filter update \cite{ENZNER20061140,kuech_state_space_2014,LuisValero2019}, denoted by \textit{KF} (cf. Eq.~\eqref{eq:kalman_nlms}). The interference power $\psi_{f,\tau}^{ZZ}$ and the error power $\psi_{f,\tau}^{EE}$ which are required for the respective updates are both estimated by recursive averaging of the magnitude-squared error $|e_{f,\tau}|^2$ with a smoothing factor of $0.5$ \cite{yang_improved_kf_with_fast_recovery}. Furthermore, the raw step-size $m^{\text{EA-NLMS}}$ and loudspeaker smoothing parameter $\lambda_U$ of the \makebox{\textit{EA-NLMS}} algorithm were chosen to $0.2$ and $0.9$, respectively, and the state transition factor and initial filter estimation variances of the Kalman filter update are chosen to $0.99$ and $1$, respectively. The filter process noise power for the Kalman filter was chosen proportionally to a temporally-smoothed average of the magnitude-squared filter coefficients as described in \cite{kuech_state_space_2014}. In addition, an additive regularization of $10^{-3}$ was added to the numerators of the \makebox{\textit{EA-NLMS}} and \textit{KF} updates (cf. Eqs. \eqref{eq:nlms_ea} and \eqref{eq:kalman_nlms}) and a minimum filter process noise power of $10^{-3}$ was chosen for the Kalman filter which drastically improved their robustness. Note that the hyperparameters have been optimized by a grid search on the test data set which favors the baseline results.

Besides the traditional step-size estimators, we consider also two oracle baselines which have knowledge about the individual signal components. On the one hand, we consider an oracle \ac{NLMS} algorithm, denoted by \makebox{\textit{Oracle-Grad-NLMS}}, which uses the true echo component $d_{f,\tau}$ instead of the microphone signal $y_{f,\tau}$ to compute the \gls{CTF} filter gradient (cf. Eqs.~\eqref{eq:err_comp_prior} and \eqref{eq:lms_update}).  
On the other hand, an interference power (IP)-informed Kalman filter approach, denoted by \makebox{\textit{Oracle-IP-KF}}, is investigated which estimates the interference power $\psi_{f,\tau}^{ZZ}$ by a recursive average of the magnitude-squared interfering signal $|z_{f,\tau}|^2 = |s_{f,\tau}+n_{f,\tau}|^2$ with a smoothing parameter of $0.5$. It is important to note that while \makebox{\textit{Oracle-Grad-NLMS}} uses the true echo signal for computing the \gls{CTF} filter update, and therefore adapts continuously, \makebox{\textit{Oracle-IP-KF}} exploits only oracle knowledge about the interference power for computing the step-size, and therefore does not adapt continuously. 
%
%
The step-size of \makebox{\textit{Oracle-Grad-NLMS}} was chosen to $0.2$ and the loudspeaker power was estimated by recursive averaging with a smoothing parameter of $\lambda_U=0.9$.
%
%
Furthermore, a regularization constant of $10^{-3}$ was added to the numerator of the update. The state transition factor, initial filter estimation variances, minimum process noise power and additive regularization of \makebox{\textit{Oracle-IP-KF}} were chosen to $0.999$, $0.1$, $10^{-4}$ and $1$, respectively. 
%
Finally, note that the results of both oracle algorithms do not pose an upper performance bound for the linear \gls{CTF} echo estimation model \eqref{eq:ctf_echo_est} as the hyperparameters are frequency independent and have been chosen equivalently for all acoustic scenes of the test data set, based on the best average performance.
%

\subsection{Performance Measures}
\label{sec:perf_measures}
We now introduce various measures to compare the performance of the different adaptation control algorithms. One of the most popular measures is given by the \gls{ERLE} \cite{enzner_acoustic_2014}
\begin{equation}
	\text{ERLE} = 10 \log_{10} \frac{\sum_{\kappa=1}^K \underline{d}_\kappa^2 }{\sum_{\kappa=1}^K (\underline{d}_\kappa - \hat{\underline{d}}_\kappa)^2}
	\label{eq:erle}
\end{equation}
which is computed by the logarithmic ratio of the time-domain echo power relative to the residual echo power after the echo canceler. It is important to note that the \gls{ERLE} definition \eqref{eq:erle} considers the true echo signal $\underline{d}_\kappa$ and not the microphone signal $\underline{y}_\kappa$ and thus can also be used to assess the echo power attenuation during double-talk. Furthermore, it is, apart from regularization constants, equivalent to the negated loss function \eqref{eq:td_erle_loss}. Due to its simple interpretability, the \gls{ERLE} is widely used in practice. Yet, it is does not consider the characteristics of the possibly distorted
desired near-end speech signal. Thus, we additionally introduce the \gls{PESQ} \cite{pesq_itu}
\begin{equation}
	\text{PESQ} = f^{\text{PESQ}}(\underline{s}_\kappa, \underline{s}_\kappa + \underline{d}_\kappa - \hat{\underline{d}}_\kappa )
	\label{eq:pesq}
\end{equation}
with the residual echo $\underline{d}_\kappa - \hat{\underline{d}}_\kappa$ representing an additive distortion. Note that we refrained from considering the background noise component $\underline{n}_{\kappa}$ in \eqref{eq:pesq} as it cannot be canceled by the echo canceler and thus only dilutes the comparability.

\subsection{Evaluation}
\label{sec:evaluation}
In the following, the different adaptation control algorithms are compared w.r.t. their performance on the $450$ acoustic scenes of the test data set (cf. Sec.~\ref{sec:exp_set}). We always provide mean and standard deviation in the format \textit{mean}/\textit{std} with the best measure of each algorithmic class being typeset in bold font. For ease of readability, we abbreviate the \gls{DNN}-based adaptation control algorithms in the results tables by \gls{BB}, \gls{NB} and \gls{HB}, respectively.

\subsubsection{Number of DNN Parameters and Runtime}
We start by comparing the algorithms in terms of \gls{DNN} parameter count, runtime $t_{\text{proc}}$ to process a single \gls{STFT} block of length $32$~ms and \gls{RTF}. Tab.~\ref{tab:runtime_memory} shows the respective measures for the considered traditional and deep learning-based adaptation control algorithms. 
%
%
\begin{table}[b]
	\vspace*{-.00cm}
	\caption{Approximate number of \gls{DNN} parameters, block processing runtime $t_{\text{proc}}$ and \gls{RTF} for a typical configuration of the different adaptation control approaches.}
	\vspace*{-.0cm}
	 \setlength{\tabcolsep}{5.0pt} 
	\begin{center}
		\begin{tabular}{c  || c c | c  c c}
			& EA-NLMS &	KF & BB-DNN & NB-DNN & HB-DNN \\
			\toprule
			\#Parameters& \textemdash & \textemdash  & $330 \cdot 10^3$ & {$\boldsymbol{50\cdot10^3} $} & $50 \cdot10^3$ \\
			$t_{\text{proc}}$ in ms & $\boldsymbol{0.11}~\hspace*{.065cm}$ & ${0.15}~\hspace*{.065cm}$ & $0.59~$ & $0.79~~$ & $0.83$ \\
			RTF & $\boldsymbol{0.014}$ & $0.018$ & $0.074$  & $0.098$ & $0.10$	
		\end{tabular} 
	\end{center}
	\label{tab:runtime_memory}
	\vspace{-.0cm} 
\end{table}
The block processing runtime $t_{\text{proc}}$ and \gls{RTF} have been measured on an \textit{Intel Xeon W-1390@2.80GHz} CPU. Note that all presented values depend on the specific algorithmic configuration, i.e., dimensions of the signal feature vectors and \gls{GRU} states. Thus, the values can only be considered as typical representatives of the respective algorithmic class. Obviously the broadband deep learning-based adaptation control (BB-DNN) exhibits significantly more parameters than the narrowband (NB-DNN) and hybrid (HB-DNN) versions which is straightforwardly explained by the larger feature, masking and \gls{GRU} dimensions. Furthermore, the narrowband and hybrid \glspl{DNN} exhibit a similar number of \gls{DNN} parameters as they differ only by an additional fully connected layer, with at most $192$ parameters, which maps from the hybrid feature vector $\boldsymbol{\xi}_{\tau}$ to the \gls{GRU} states (cf. Fig.~\ref{fig:nb_bb_hybrid_dnn_vis}). Considering the runtime, the \gls{DNN}-based algorithms are approximately four to seven times slower than their traditional counterparts with the hybrid method exhibiting the highest runtime. Yet, it should be noted that the runtime of all algorithms could easily be reduced by subsampling the step-size inference, i.e., keeping it constant for several successive frames. This is in particular attractive for an \gls{STFT} with a quarter-frameshift as considered in this paper. Finally, it might seem counterintuitive that the broadband method has a lower processing runtime than the narrowband/hybrid methods, despite exhibiting more parameters. Yet, this is straightforwardly explained by considering that the broadband approach requires only a single \gls{DNN} inference to jointly estimate the step-sizes for all frequency bands while the narrowband and hybrid methods execute individual \gls{DNN} inferences per frequency band. 

\subsubsection{Loss Functions}
We now investigate the different loss functions for \gls{DNN} training that have been introduced in Sec.~\ref{sec:loss_func}. Tab.~\ref{tab:loss_func_comp_tab} shows the performance evaluation for the broadband, narrowband and hybrid \gls{DNN}-based adaptation control algorithms after being trained with the different loss functions \eqref{eq:fd_mse_loss} - \eqref{eq:td_erle_loss}. 
\begin{table}[b]
	\vspace*{-.00cm}
	\caption{Loss function comparison for broadband (BB), narrowband (NB) and hybrid (HB) \gls{DNN}-based adaptation control algorithms with the absolute \gls{STFT}-domain loudspeaker and microphone signals as feature vectors.}
	\vspace*{-.0cm}
	\setlength{\tabcolsep}{8.0pt}
	\begin{center}
		\begin{tabular}{c c || c  c}
			Loss (Eq.)  & Structure & ERLE & PESQ \\
			\midrule
	\multirow{3}{*}{	FD-MSE \eqref{eq:fd_mse_loss}}   
						& 	BB-DNN  & $12.79/3.1$ &  $1.93/0.8$ \\
						& 	NB-DNN  & $14.54/3.3$ &  $\textbf{2.11}/0.8$ \\
						& 	HB-DNN  &$\textbf{14.66}/3.3$ &  $\textbf{2.11}/0.8$ \\
						\midrule
			\multirow{3}{*}{	TD-MSE \eqref{eq:td_mse_loss}}   
			& 	BB-DNN  & $12.84/3.2$ &  $1.92/0.8$ \\
			& 	NB-DNN  & $14.48/3.3$ &  $2.10/0.8$ \\
			& 	HB-DNN  & $\textbf{14.61}/3.3$ &  $\textbf{2.11}/0.8$ \\
			\midrule
			\multirow{2}{*}{	FD-ERLE \eqref{eq:fd_erle_loss}}   
			& 	BB-DNN  & $12.78/3.5$ &  $1.94/0.8$ \\
			& 	NB-DNN  & $14.51/3.4$ &  $2.10/0.8$ \\
			& 	HB-DNN  & $\textbf{14.70}/3.4$ &  $\textbf{2.12}/0.8$ \\
			\midrule
						\multirow{3}{*}{	TD-ERLE \eqref{eq:td_erle_loss}}   
			& 	BB-DNN  & $12.99/3.3$ &  $1.94/0.8$ \\
			& 	NB-DNN  & ${14.71}/3.4$ &  $\textbf{2.12}/0.8$ \\
			& 	HB-DNN  & $\textbf{14.85}/3.4$ &  $\textbf{2.12}/0.8$ \\
		\midrule\midrule
		\multicolumn{2}{c||}{{EA}-NLMS} & $10.16/3.3$ &  $1.81/0.7$ \\
     \multicolumn{2}{c||}{KF} & $\textbf{11.99}\hspace*{.01cm}/3.4$ &  $\textbf{1.96}/0.7$ \\
    \midrule\midrule
\multicolumn{2}{c||}{Oracle-IP-KF} & $14.90/3.6$ &  $2.16/0.8$ \\
		\multicolumn{2}{c||}{Oracle-Grad-NLMS} & $\textbf{15.17}/3.7$ &  $\textbf{2.20}/0.9$ \\
		\end{tabular} 
	\end{center}
	\label{tab:loss_func_comp_tab}
\end{table}
As feature vectors we considered the absolute \gls{STFT}-domain loudspeaker and microphone signals and the \gls{DNN} output layer was chosen according to \eqref{eq:dnn_ea_nlms}. 
We conclude from Tab.~\ref{tab:loss_func_comp_tab} that all deep learning-based step-size estimators (cf.~rows~$1-4$) outperform the traditional error power-aware \gls{NLMS} and Kalman filter baselines (cf. EA-NLMS and KF) in terms of \gls{ERLE}. Among the \gls{DNN}-based step-size estimators, the narrowband and hybrid methods outperform their broadband counterparts by approximately $1.7$~dB \gls{ERLE} and $0.17$ \gls{PESQ} on average. 
It might seem counterintuitive that the narrowband and hybrid models perform better than the broadband approach despite having much fewer \gls{DNN} parameters (cf. Tab.~\ref{tab:runtime_memory}). Yet, it should be noted again that both the narrowband and the hybrid algorithms use individual \glspl{DNN}, with different internal states, per frequency band. Considering the various loss functions, we observe only minor differences with small benefits for the TD-ERLE cost function \eqref{eq:td_erle_loss}. Thus, we conclude that the different deep learning-based adaptation control algorithms are highly robust w.r.t. the cost function choice and choose the TD-ERLE cost function \eqref{eq:td_erle_loss} for the following experiments.

\subsubsection{DNN Feature Vectors}
%
In this section, the influence of the feature vector choice (cf. Sec.~\ref{sec:feat_vec}) on the echo cancellation performance is investigated. We first examine the different transformations for mapping from the complex-valued signals to real-valued features. Tab.~\ref{tab:feat_trans_res_tab} shows the average performance measures of the narrowband and broadband \gls{DNN}-based step-size estimators for the real- and imaginary (cf. Eq.~\eqref{eq:re_im_feat}), magnitude (cf. Eq.~\eqref{eq:mag_feat}) and logarithmic magnitude (cf. Eq.~\eqref{eq:log_abs_feat_trans}) transformations of the complex-valued loudspeaker and microphone signals.
\begin{table}[t]
	\vspace*{-.00cm}
	\caption{Comparison of different feature transformations (cf. Eqs.~\eqref{eq:re_im_feat} - \eqref{eq:log_abs_feat_trans}) for broadband (BB) and narrowband (NB) deep learning-based adaptation control.}
	\vspace*{-.0cm}
	\begin{center}
		\begin{tabular}{c |c || c  c}
		Structure  &Feature Mapping & ERLE  &PESQ  \\
			\midrule
			\multirow{3}{*}{BB-DNN}   
			& $ \left( \mathcal{R}\{\cdot \} ,  \mathcal{I}\{\cdot \}\right)$ &   $11.59/3.2$ &  $1.85/0.7$ \\
			& $\left| \cdot \right|$ & $\textbf{12.99}/3.3$ & $\textbf{1.94}/0.8$ \\
			& $ \log_{10} \left( 	\left| \cdot \right|  + \delta^\text{FEAT} \right)$ & $12.88/3.4$ &  $\textbf{1.94}/0.8$ \\
			\midrule
			\multirow{3}{*}{	NB-DNN}   
			& $ \left( \mathcal{R}\{\cdot \} ,  \mathcal{I}\{\cdot \}\right)$ &   {$14.67/3.4$} &  {$\textbf{2.12}/0.8$} \\
			& $\left| \cdot \right|$ & $\textbf{14.71}/3.4$ &  $\textbf{2.12}/0.8$ \\
			& $ \log_{10} \left( 	\left| \cdot \right|  + \delta^\text{FEAT} \right)$ &$14.66/3.4$ &  $\textbf{2.12}/0.8$ \\
		\end{tabular} 
	\end{center}
	\label{tab:feat_trans_res_tab}
	\vspace{-.2cm} 
\end{table}
We conclude that all feature transformations perform similar for the narrowband method. In contrast, a clear benefit for the magnitude-based features in comparison to the real and imaginary features can be observed for the broadband method. This might be surprising as the latter contain the most information and the magnitude could be computed from the real and imaginary components. Yet, this might require more powerful \glspl{DNN} architectures. It indicates also that the phase information is less relevant for adaptation control which is also suggested by many traditional step-size estimators (cf. Eqs.~\eqref{eq:nlms_bin} - \eqref{eq:nlms_vsss}).
%
%
In addition, it should be considered that the feature dimension of the real and imaginary feature vector is twice the dimension of the magnitude-based feature vectors. As the simple magnitude transformation tends to perform best, it is chosen as basis for the subsequent investigations.
\begin{table}[t]
	\vspace*{-.00cm}
	\caption{Comparison of different magnitude-based signal features for broadband (BB), narrowband (NB) and hybrid (HB) \gls{DNN}-based adaptation control algorithms.}
	\vspace*{-.0cm}
	\setlength{\tabcolsep}{4.5pt}
	\begin{center}
		\begin{tabular}{c |c| c |c |c |c |c |c || c  c}
		\multirow{2}{*}{Structure}  &\multicolumn{7}{c||}{Features} & \multirow{2}{*}{ERLE}  & \multirow{2}{*}{PESQ}  \\
		  & $u_{f}$ & $y_{f}$ & $e_{f}$ & $\hat{d}_{f}$ & $\bar{y}$ &  $\bar{e}$ &  $\bar{\hat{d}}$ & & \\
			\midrule
			\multirow{5}{*}{	BB-DNN}   
					& \checkmark & \checkmark &  &  &  &  &  & {$12.99/3.3$} &  {$1.94/0.8$} \\
					& \checkmark &  & \checkmark &  &  &  &  & {$13.49/3.5$} &  {$1.99/0.8$} \\
					&  & \checkmark &  & \checkmark &  &  &  & {$13.35/3.5$} &  {$1.98/0.8$} \\
					& \checkmark  & \checkmark & \checkmark &  &  &  &  & {$13.71/3.5$} &  {$2.00/0.8$} \\
					& \checkmark & \checkmark & \checkmark & \checkmark &  &  &  & {$\textbf{14.04}/3.4$} & {$\textbf{2.03}/0.8$} \\
			\midrule
			\multirow{5}{*}{	NB-DNN}   
					& \checkmark & \checkmark &  &  &  &  &  & {$14.71/3.4$} &  {$2.12/0.8$} \\
& \checkmark &  & \checkmark &  &  &  &  & {$14.60/3.5$} &  {$2.13/0.8$} \\
&  & \checkmark &  & \checkmark &  &  &  & {$14.71/3.4$} &  {$2.12/0.8$} \\
& \checkmark  & \checkmark & \checkmark &  &  &  &  & {$15.10/3.5$} &  {$2.15/0.8$} \\
& \checkmark & \checkmark & \checkmark & \checkmark &  &  &  & {$\textbf{15.17}/3.5$} & {$\textbf{2.16}/0.8$} \\
			\midrule
			\multirow{4}{*}{	HB-DNN}   
					& \checkmark & \checkmark &  &  & \checkmark &  &  & $14.85/3.4$ &  {$2.12/0.8$} \\
& \checkmark &  & \checkmark &  &  & \checkmark &  & {$14.80/3.6$} &  {$2.13/0.8$} \\
& \checkmark  & \checkmark & \checkmark &  & \checkmark & \checkmark &  & {${15.35}/3.5$} &  {${2.16}/0.8$} \\
& \checkmark  & \checkmark & \checkmark & \checkmark & \checkmark & \checkmark & \checkmark & {$\textbf{15.60}/3.5$} &  {$\textbf{2.18}/0.8$} \\
			\midrule\midrule
\multicolumn{8}{c||}{{EA}-NLMS} & $10.16/3.3$ &  $1.81/0.7$ \\
			\multicolumn{8}{c||}{KF} & $\textbf{11.99}\hspace*{.01cm}/3.4$ &  $\textbf{1.96}/0.7$ \\
\midrule\midrule
\multicolumn{8}{c||}{Oracle-IP-KF} & $14.90/3.6$ &  $2.16/0.8$ \\
\multicolumn{8}{c||}{Oracle-Grad-NLMS} & $\textbf{15.17}/3.7$ &  $\textbf{2.20}/0.9$ \\
		\end{tabular} 
	\end{center}
	\label{tab:feat_res_tab}
\vspace{-.2cm} 
\end{table}

We now examine the effect of different signal choices for computing the magnitude-based features. The average performance measures of the accordingly trained broadband, narrowband and hybrid \gls{DNN}-based adaptation control algorithms is shown in Tab.~\ref{tab:feat_res_tab}.
We observe that in general increasing the feature set improves the adaptation control performance. In particular, the joint usage of the microphone and the error signal seems to be beneficial. This might be explained by the importance of the convergence state information of the echo canceler that is difficult to extract from the respective individual signals. We finally observe that the best-performing \gls{DNN}-based step-size estimator outperforms the oracle \gls{NLMS} method in terms of \gls{ERLE} and almost attains its \gls{PESQ} values. 
Yet, it should be noted again that both oracle algorithms assume equivalent (frequency independent) hyperparameters for all acoustic scenes of the test data set.

\subsubsection{Output Layer Design}
%
Finally, we investigate different output layer designs of the \gls{DNN} (Sec.~\ref{sec:out_layer_design}). In particular, we examine the effect of omitting the adaptive error power normalization in Eq.~\eqref{eq:dnn_ea_nlms} by setting $m^{\text{DNN-e}}_{f,\tau}=0$ or a non-adaptive selection by setting $m^{\text{DNN-e}}_{f,\tau}=1$. Furthermore, we evaluate a frequency-independent selection of the step-size masks $m^{\text{DNN-$\mu$}}_{f,\tau}$ and $m^{\text{DNN-e}}_{f,\tau}$, respectively, for the broadband method, i.e., an identical choice for all frequency bands. As feature vectors we use the magnitude loudspeaker, microphone and error signals which showed good performance according to Tab.~\ref{tab:feat_res_tab}. The respective \gls{ERLE} and \gls{PESQ} results are shown in Tab.~\ref{tab:out_layer_des}. 
We conclude that frequency-selective step-size masks significantly outperform a frequency-independent choice (cf. BB-DNN) which can be explained by the non-whiteness which is typical for speech and audio signals, both as desired signals and as interference. Furthermore, a general error power normalization within the output layer of the broadband approach, either controlled by a \gls{DNN} (cf. $m^{\text{DNN-e}}_{f,\tau}$$=${selective}) or a static choice (cf. $m^{\text{DNN-e}}_{f,\tau}$$=$$1$), is beneficial in comparison to a simple \gls{DNN}-controlled \ac{NLMS}~(cf.  $m^{\text{DNN-e}}_{f,\tau}$$=$$0$). Yet, the \gls{DNN} seems to infer values close to one as {$m^{\text{DNN-e}}_{f,\tau}$$=${selective}} and $m^{\text{DNN-e}}_{f,\tau}$$=${$1$} show a similar performance. In contrast, no clear benefit of the adaptive error-power normalization within the narrowband and hybrid approaches can be observed. This might be a result of the step-size inference by individual \glspl{DNN} per frequency band which can handle for the considered architectures more challenging situations in comparison to the broadband approach. Thus, a more conservative, i.e., smaller, step-size choice by error power normalization is not required.
\begin{table}[t]
	\vspace*{-.00cm}
	\caption{Comparison of different output layer designs (cf. Eq.~\eqref{eq:dnn_ea_nlms}) for \gls{DNN}-based step-size estimation. While \textit{selective} approaches estimate a different mask for each frequency band, \textit{non-selective} methods estimate a frequency-independent mask which is equivalently applied to all frequency bands.}
	\vspace*{-.0cm}
	\setlength{\tabcolsep}{4.5pt}
	\begin{center}
		\begin{tabular}{c |c| c  || c  c}
			\multirow{2}{*}{Structure}  &\multicolumn{2}{c||}{Output layer} & \multirow{2}{*}{ERLE}  & \multirow{2}{*}{PESQ}  \\
			& $m^{\text{DNN-$\mu$}}_{f,\tau}$ & $m^{\text{DNN-e}}_{f,\tau}$ & & \\
			\midrule
			\multirow{6}{*}{	BB-DNN}   
			& Selective &  Selective  &     $\textbf{13.71}/3.5$ &  ${2.00}/0.8$ \\
			& Selective & $0$   &   {$13.03/3.5$} &  $1.95/0.8$ \\
			& Selective & $1$  &    $13.70/3.4$ &  $\textbf{2.01}/0.8$ \\
			& Non-Selective &  Non-Selective  &  {$12.97/3.4$} &  {$1.90/0.7$} \\
			&  Non-Selective & $0$   &   $11.45/3.8$ &  $1.76/0.7$ \\
			&  Non-Selective & $1$  &   $12.87/3.3$ &  $1.89/0.7$ \\
			\midrule
			\multirow{3}{*}{	NB-DNN}   
			& Selective &  Selective  &  $\textbf{15.10}/3.5$ &  $\textbf{2.15}/0.8$ \\
& Selective & $0$   &   {$15.07/3.5$} &  {$\textbf{2.15}/0.8$} \\
& Selective & $1$  &    {$14.92/3.4$} &  {$2.14/0.8$} \\
			\midrule
			\multirow{3}{*}{	HB-DNN}   
			& Selective &  Selective  &    $\textbf{15.35}/3.5$ &  $\textbf{2.16}/0.8$ \\
& Selective & $0$   &   {$15.30/3.5$} &  {$\textbf{2.16}/0.8$} \\
& Selective & $1$  &    $15.13/3.5$ &  $\textbf{2.16}/0.8$ \\
		\end{tabular} 
	\end{center}
	\label{tab:out_layer_des}
	\vspace{-.5cm} 
\end{table}

\subsection{Visual Interpretation}
\label{sec:exp_step_size_vis}
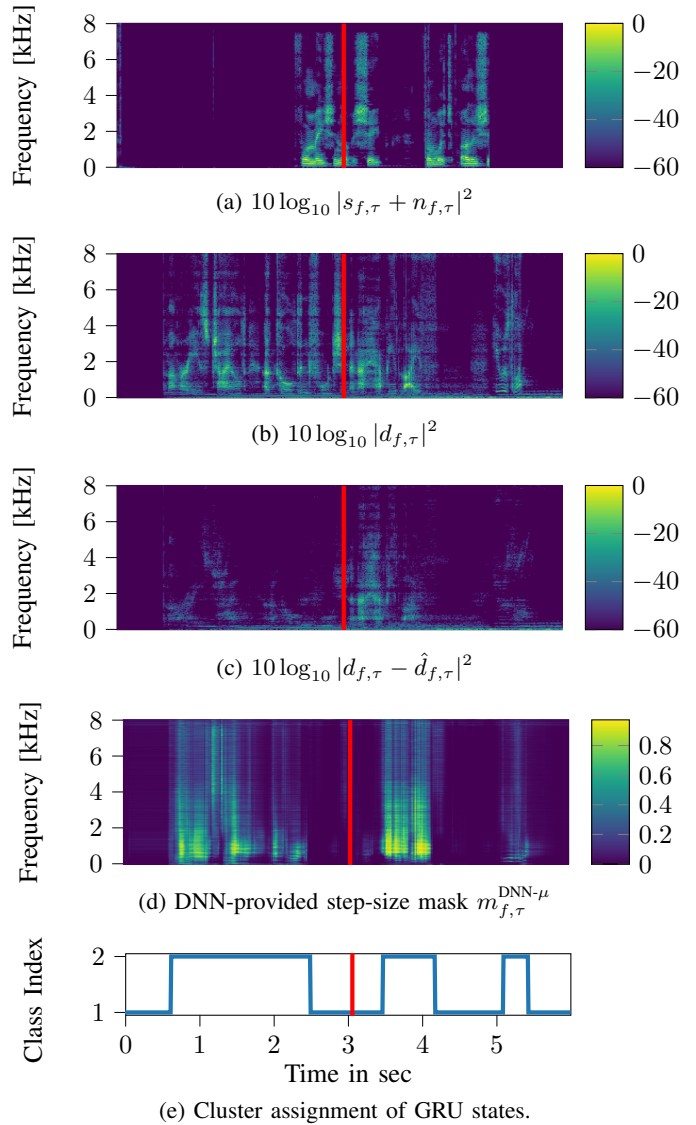
\begin{figure}[t]
	\centering
	 \begin{subfigure}[b]{0.5\textwidth}
		\centering
\begin{tikzpicture}

\definecolor{darkgray176}{RGB}{176,176,176}

\begin{axis}[
	width=7.5cm, height=3.5cm,
	colorbar,
	colorbar style={ylabel={}},
	colormap/viridis,
	point meta max=0,
point meta min=-60,
tick align=outside,
tick pos=left,
x grid style={darkgray176},
xmajorticks=false,
xmin=2, xmax=7.996003996004,
xtick style={color=black},
y grid style={darkgray176},
ylabel={Frequency [kHz]},
ymin=-0.015625, ymax=8.015625,
ytick style={color=black}
]
\addplot graphics [includegraphics cmd=\pgfimage,xmin=-0.003996003996004, xmax=7.996003996004, ymin=-0.015625, ymax=8.015625] {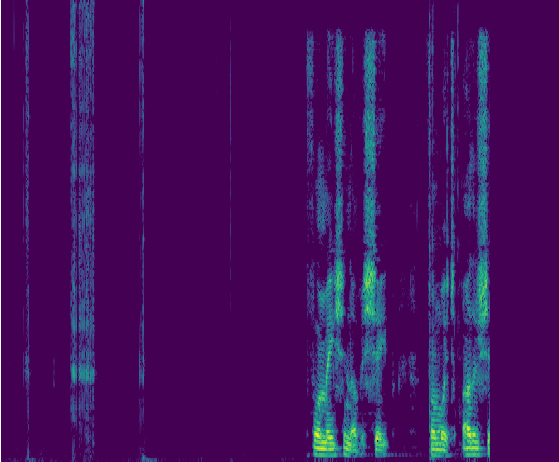};
\addplot [ultra thick, red]
table {%
    5.05318737030029 -8
	5.05318737030029 8
};
\end{axis}

\end{tikzpicture}
		\vspace*{-.3cm}
		\caption{$10 \log_{10} |s_{f,\tau} + n_{f,\tau}|^2$}
		\label{fig:step_size_vis_1}
	\end{subfigure}
	\vspace*{-.3cm}
	
	\begin{subfigure}[b]{0.5\textwidth}
		\centering
\begin{tikzpicture}

\definecolor{darkgray176}{RGB}{176,176,176}

\begin{axis}[
	width=7.5cm, height=3.5cm,
colorbar,
colorbar style={ylabel={}},
colormap/viridis,
point meta max=0,
point meta min=-60,
tick align=outside,
tick pos=left,
x grid style={darkgray176},
xmajorticks=false,
xmin=2, xmax=7.996003996004,
xtick style={color=black},
y grid style={darkgray176},
ylabel={Frequency [kHz]},
ymin=-0.015625, ymax=8.015625,
ytick style={color=black}
]
\addplot graphics [includegraphics cmd=\pgfimage,xmin=-0.003996003996004, xmax=7.996003996004, ymin=-0.015625, ymax=8.015625] {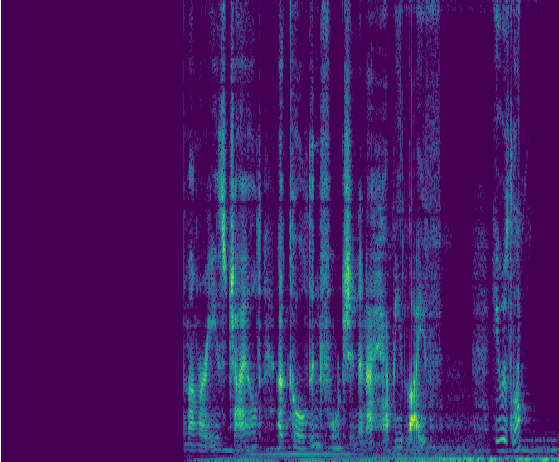};
\addplot [ultra thick, red]
table {%
	5.05318737030029 -8
	5.05318737030029 8
};
\end{axis}

\end{tikzpicture}
			\vspace*{-.7cm}
					\caption{$10 \log_{10} |d_{f,\tau} |^2$}
	\end{subfigure}
	\vspace*{-.3cm}
		
			\begin{subfigure}[b]{0.5\textwidth}
			\centering
\begin{tikzpicture}

\definecolor{darkgray176}{RGB}{176,176,176}

\begin{axis}[
	width=7.5cm, height=3.5cm,
	colorbar,
	colorbar style={ylabel={}},
	colormap/viridis,
	point meta max=0,
point meta min=-60,
tick align=outside,
tick pos=left,
x grid style={darkgray176},
xmajorticks=false,
xmin=2, xmax=7.9960039960,
xtick style={color=black},
y grid style={darkgray176},
ylabel={Frequency [kHz]},
ymin=-0.015625, ymax=8.015625,
ytick style={color=black}
]
\addplot graphics [includegraphics cmd=\pgfimage,xmin=-0.003996003996004, xmax=7.996003996004, ymin=-0.015625, ymax=8.015625] {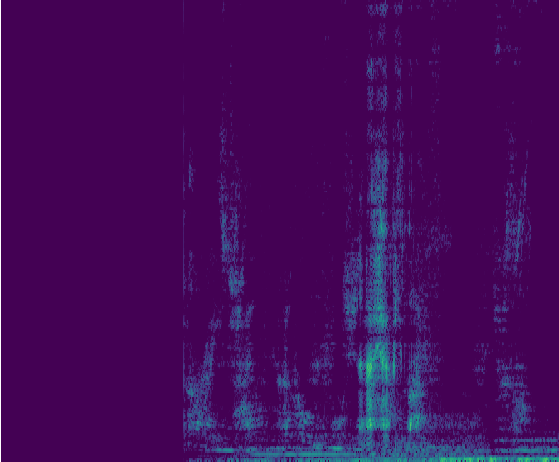};
\addplot [ultra thick, red]
table {%
	5.05318737030029 -8
	5.05318737030029 8
};
\end{axis}

\end{tikzpicture}
			\vspace*{-.7cm}
								\caption{$10 \log_{10} |d_{f,\tau} - \hat{d}_{f,\tau} |^2$}
		\end{subfigure}
	\vspace*{-.3cm}

\begin{subfigure}[b]{0.5\textwidth}
	\centering
		\hspace*{-.1cm}
\begin{tikzpicture}

\definecolor{darkgray176}{RGB}{176,176,176}

\begin{axis}[
	width=7.5cm, height=3.5cm,
		colorbar,
		colorbar style={ylabel={}},
		colormap/viridis,
		point meta max=0.973592936992645,
		point meta min=4.04311492729903e-07,
tick align=outside,
tick pos=left,
x grid style={darkgray176},
xmajorticks=false,
xmin=2, xmax=7.996003996004,
xtick style={color=black},
y grid style={darkgray176},
ylabel={Frequency [kHz]},
ymin=-0.015625, ymax=8.015625,
ytick style={color=black}
]
\addplot graphics [includegraphics cmd=\pgfimage,xmin=-0.003996003996004, xmax=7.996003996004, ymin=-0.015625, ymax=8.015625] {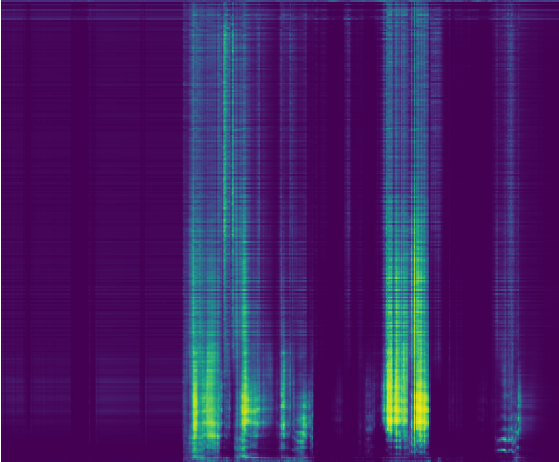};

\addplot [ultra thick, red]
table {%
	5.05318737030029 -8
	5.05318737030029 8
};
\end{axis}

\end{tikzpicture}
\vspace*{-.7cm}

	\caption{\gls{DNN}-provided step-size mask $m^{\text{DNN-$\mu$}}_{f,\tau}$} 
	\label{fig:step_size_vis_3}
\end{subfigure}
	\vspace*{-.3cm}

\begin{subfigure}[b]{0.5\textwidth}
	\centering

    \hspace*{-.15cm}
\begin{tikzpicture}

\definecolor{darkgray176}{RGB}{176,176,176}
\definecolor{steelblue31119180}{RGB}{31,119,180}

\begin{axis}[
	width=7.5cm, height=2.4cm,
tick align=outside,
tick pos=left,
xlabel={Time in sec},
x grid style={darkgray176},
xmin=0, xmax=5.996003996004,
xtick style={color=black},
y grid style={darkgray176},
ymin=0.95, ymax=2.05,
ytick={1, 2},
ylabel={Class Index},
y label style={xshift=-0.2cm},
ytick style={color=black}
]
\addplot [ultra thick, steelblue31119180]
table {%
-2 1
0.6053946018219 1
0.61338663101196 2
2.48351669311523 2
2.49150848388672 1
3.45854139328003 1
3.46653366088867 2
4.16183805465698 2
4.16983032226562 1
5.08091926574707 1
5.08891105651855 2
5.41658353805542 2
5.4245753288269 1
5.99200820922852 1
};
\addplot [ultra thick, red]
table {%
3.05318737030029 -4
3.05318737030029 4
};
\end{axis}

\end{tikzpicture}\hspace*{1.1cm}
			\vspace*{-.1cm}
	\caption{Cluster assignment of \gls{GRU} states.}
	\label{fig:step_size_vis_4}
\end{subfigure}

	\caption{Logarithmic interference, echo and residual echo power spectrograms in comparison to the \gls{DNN}-provided step-size mask $m^{\text{DNN-$\mu$}}_{f,\tau}$ with the red line indicating an echo path change. In Subfig.~\ref{fig:step_size_vis_4}, the frame-based assignment of the internal \gls{GRU} states of the \gls{DNN} to two different classes/clusters is shown.}
	\label{fig:step_size_vis}
 \vspace*{-.1cm}
\end{figure}

After comparing the various approaches numerically, we will now visually interpret the \gls{DNN}-based step-size estimates. For this we show in  Fig.~\ref{fig:step_size_vis} the logarithmic interference, i.e., summed near-end speech and background noise, echo and residual echo signal power spectrograms relative to the \gls{DNN}-provided step-size mask $m^{\text{DNN-$\mu$}}_{f,\tau}$ for a specific sequence. Note that around $3.1$ s an echo path change occurs which is indicated by the vertical red lines. A broadband approach without any error-power normalization, i.e., $m^{\text{DNN-e}}_{f,\tau}$ $=$ $0$, and only a single \gls{GRU} layer with $128$ internal states was used to create the results. The stacked magnitude loudspeaker, microphone and error signals were used as feature vectors. To gain more insights into what the \gls{DNN} actually learns, we additionally try to find similar \gls{GRU} state vectors across temporally different frames \cite{briegleb2023localizing}. For this we cluster the respective \gls{GRU} state vectors of the sequence by agglomerative clustering with the city block metric into two different clusters \cite{scikit-learn}. Subfig.~\ref{fig:step_size_vis_4} shows the frame-based assignment of the internal \gls{GRU} state vectors to the two different classes. We observe from the cluster assignment results in Subfig.~\ref{fig:step_size_vis_4} that the first class indicates error signal frames which are dominated by interference, i.e., near-end speech or background noise (or no echo activity), whereas the second cluster indicates error signal frames which are dominated by residual echo. 
Thus, we conclude that the \gls{GRU} states encode important information about the interference activity which points to the interpretation that the \gls{DNN} includes a refined double-talk detector.
%
%
%
The step-size vectors (columns in Subfig.~\ref{fig:step_size_vis_3}) which are computed from the \gls{GRU} states of the second cluster exhibit a clear \textit{low-pass} structure, i.e., high-values in the low frequency range, which is reasonable considering the speech loudspeaker signal. 
%
Of particular interest is the step-size selection after the echo path change: While there is no adaptation during the initial double-talk period from $3.1~\text{s}-3.5~\text{s}$, the subsequent decreasing interference power results in high step-sizes and thus rapid filter updates.

\section{Conclusion}
\label{sec:conclusion}
In this paper, we investigated various \gls{DNN}-based adaptation control methods for frequency domain acoustic echo cancellation. In particular, we compared broadband and narrowband step-size inference approaches and proposed a hybrid method which combines the best out of both worlds. Our results show significant performance improvements of deep learning-based step-size estimation relative to traditional approaches, in particular for dynamic acoustic scenes. Furthermore, we provide a general view on different signal feature vectors, cost functions and \gls{DNN} output layers which yields novel insights and guidelines for the design of other deep learning-based adaptation control algorithms. As future work, our generic step-size adaptation approach can be applied to unsupervised system identification applications, e.g., relative transfer function estimation, and to the joint control of various algorithmic components of a multichannel hands-free speech interface, e.g., echo canceler, beamformer and spectral postfilter, by a single \gls{DNN} as outlined in \cite{dnn_aec_bf_dpf}. 


\bibliographystyle{ieeetran}
\bibliography{IEEEabrv,./references}

\end{document}